\documentclass[10pt]{article}
\pdfoutput=1
\usepackage{geometry}
\usepackage{amsfonts}
\usepackage{amsmath}
\usepackage{amssymb}
\usepackage{mathptmx}
\usepackage{graphicx}
\usepackage{hyperref}
\usepackage{cancel}
\usepackage{textcomp}
\usepackage{bm}
\usepackage{times}
\usepackage{epsfig}
\usepackage{xcolor}
\usepackage{colortbl}
\usepackage{slashed}
\usepackage{booktabs}
\usepackage{multirow}
\usepackage{latexsym}
\usepackage{verbatim}
\definecolor{mygray}{gray}{.95}
\usepackage[title]{appendix}
\allowdisplaybreaks[4]

\newcommand{\diag}{\rm diag}
\newcommand{\Tr}{\rm Tr}

\newcommand{\calO}{{\cal O}}

\newcommand{\TeV}{\rm TeV}
\newcommand{\GeV}{\rm GeV}

\begin{document}
\baselineskip=16pt

\pagenumbering{arabic}

\vspace{1.0cm}

\begin{center}
{\Large\sf Effective field theory approach to lepton number violating decays $K^\pm\rightarrow \pi^\mp l^{\pm}l^{\pm}$: short-distance contribution}
\\[10pt]
\vspace{.5 cm}

{Yi Liao~$^{a,c}$\footnote{liaoy@nankai.edu.cn},~
Xiao-Dong Ma~$^{a,b}$\footnote{maxid@mail.nankai.edu.cn},~
Hao-Lin Wang~$^{a}$\footnote{whaolin@mail.nankai.edu.cn}
}

{
$^a$~School of Physics, Nankai University, Tianjin 300071, China
\\
$^b$ Department of Physics, National Taiwan University, Taipei 10617, Taiwan,
\\
$^c$ Center for High Energy Physics, Peking University, Beijing 100871, China}

\vspace{2.0ex}

{\bf Abstract}
\end{center}

This is the first paper of our systematic efforts on lepton number violating (LNV) hadronic decays in the effective field theory approach. These decays provide information complementary to popular nuclear neutrinoless double-$\beta$ ($0\nu\beta\beta$) decay in that they can probe LNV interactions involving heavier quarks and charged leptons. We may call them hadronic $0\nu\beta\beta$ decays in short, though $\beta$ refers to all charged leptons. In this work we investigate the decays $K^\pm\rightarrow\pi^\mp l^{\pm}l^{\pm}$ that arise from short-distance or contact interactions involving four quark fields and two charged lepton fields, which have canonical dimension nine (dim-9) at leading order in low energy effective field theory (LEFT). We make a complete analysis on the basis of all dim-9 operators that violate lepton number by two units, and compute their one-loop QCD renormalization effects. We match these effective interactions in LEFT to those in chiral perturbation theory ($\chi$PT) for pseudoscalar mesons, and determine the resulting hadronic low energy constants (LECs) by chiral symmetry and lattice results in the literature. The obtained decay rate is general in that all physics at and above the electroweak scale is completely parameterized by the relevant Wilson coefficients in LEFT and hadronic LECs in $\chi$PT. Assuming the standard model effective field theory (SMEFT) is the appropriate effective field theory between some new physics scale and the electroweak scale, we match our LEFT results to SMEFT whose leading effective interactions arise from LNV dim-7 operators. This connection to SMEFT simplifies significantly the interaction structures entering in the kaon decays, and we employ the current experimental bounds to set constraints on the relevant Wilson coefficients in SMEFT.

\newpage

%%%%%
\section{Introduction}

Neutrino oscillation experiments have confirmed that neutrinos have mass and their weak interactions with charged leptons mix lepton flavors. However, the origin of mass and the nature of neutrinos are still unclear. Being neutral, neutrinos could be Majorana particles and would thus violate lepton number conservation that arises as an accidental symmetry in the standard model (SM). This issue can be addressed both at high energy colliders where a typical signal for lepton number violation would be pair production of like-sign charged leptons from new heavy particles, and in high intensity experiments where the most popular so far is the nuclear neutrinoless double $\beta$ ($0\nu\beta\beta$) decay appearing as a low energy manifestation of new physics at a high scale. The two types of experiments are necessary and complementary in searching for imprints of the Majorana nature of neutrinos.

Starting from this work we will make a systematic study on lepton number violating (LNV) decays of the mesons and the $\tau$ lepton. The motivations for the efforts are evident. There are rich data on the LNV decays of the charged mesons such as $K^\pm,~D^\pm,~D^\pm_s,~B^\pm$ and the $\tau$ lepton from experiments such as LHCb, Babar, Belle, etc~\cite{Amhis:2016xyh,CortinaGil:2019dnd,Appel:2000tc,Aaij:2013sua,
Rubin:2010cq,Lees:2011hb,Kodama:1995ia,Aaij:2014aba,BABAR:2012aa,
Lees:2013gdj,Aaij:2011ex,Aaij:2012zr,Seon:2011ni,Miyazaki:2012mx}, and the bounds on some of the decays are expected to be considerably improved in proposed or upgraded experiments. This is particularly relevant considering the null results in current experiments on nuclear $0\nu\beta\beta$ decays; for reviews, see, e.g., refs.~\cite{DellOro:2016tmg,Dolinski:2019nrj}. From the theoretical point of view the above decays involve heavier quarks and charged leptons that do not appear in nuclear $0\nu\beta\beta$ decays, and thus can at least provide new information on lepton number violation that cannot be extracted from nuclear $0\nu\beta\beta$ decays. Since heavy quarks and leptons may have enhanced interactions with new particles compared to light quarks and the electron, this might also be the case with LNV processes at low energy, although we are aware that the data samples in meson decays are generally much less than in nuclear $0\nu\beta\beta$ decays. Another advantage is that one avoids uncertainties associated with nuclear physics in nuclear $0\nu\beta\beta$ decays and that for kaon and $B$ mesons and the $\tau$ lepton we can employ well-established chiral perturbation theory or heavy quark effective theory whose errors can be systematically estimated.

\begin{figure}
\centering
\includegraphics[width=12cm]{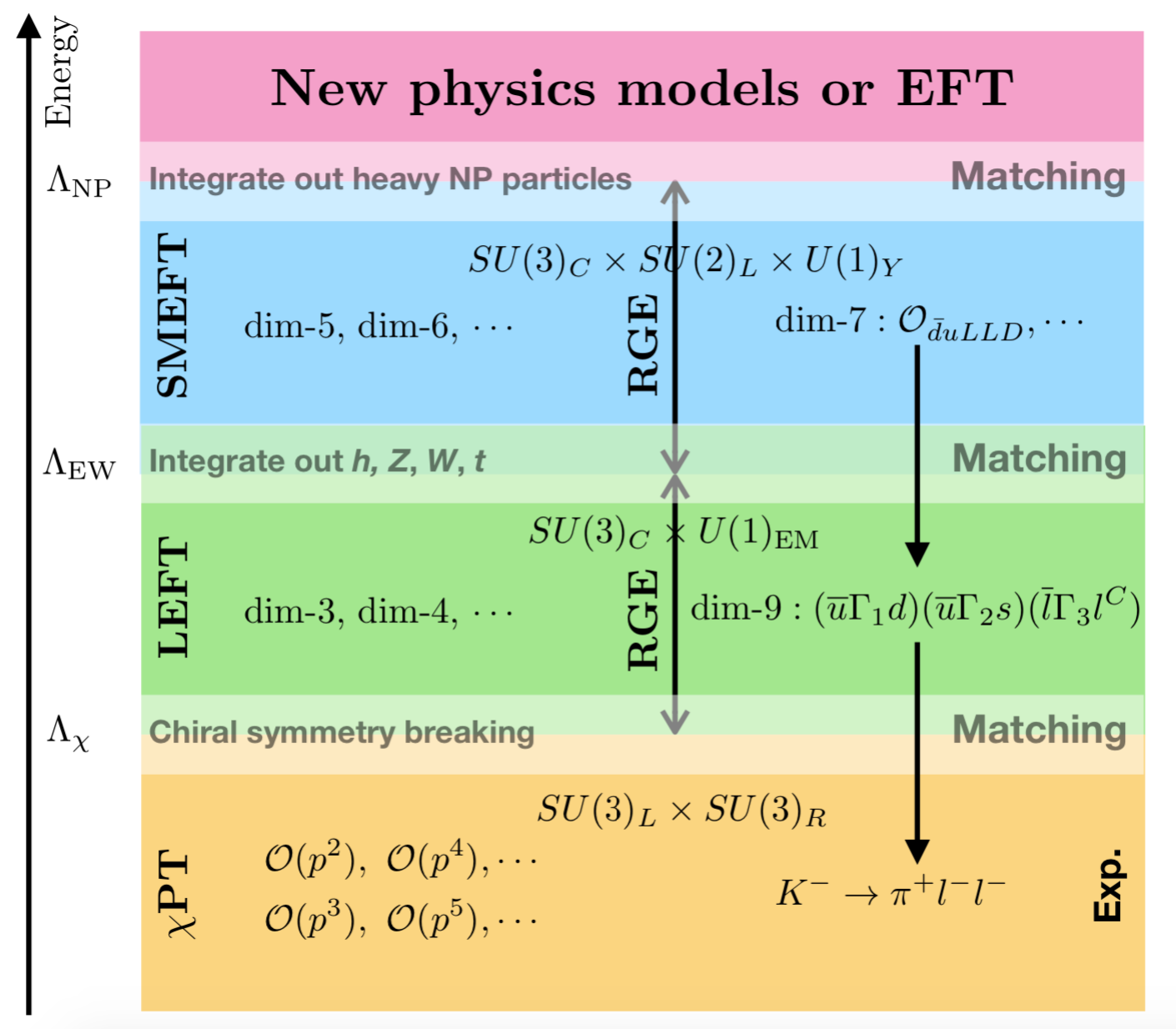}
\caption{Flow chart for a general EFT study is exemplified by the decay $K^-\rightarrow\pi^+l^-l^-$ from SMEFT, through LEFT, and to $\chi$PT, with a sequence of matching calculation and renormalization group equations (RGEs).}
\label{fig1}
\end{figure}

In this work we will investigate the decays $K^\pm\rightarrow\pi^\mp l^\pm l^\pm$ ($l=e,~\mu$) in the framework of effective field theory (EFT). Our approach is pictorially explained in figure~\ref{fig1} which shows the series of EFTs relevant to the decays. We start with the low energy effective field theory (LEFT) for quarks (excluding the top) and leptons that enjoys the QCD and QED gauge symmetries $SU(3)_C\times U(1)_\textrm{EM}$~\cite{Jenkins:2017jig,Jenkins:2017dyc}. There are then two types of contributions to the decays at the quark level: one is long-distance and the other is short-distance. We will concentrate in what follows on the short-distance or contact contribution, and reserve the long-distance part for a separate publication~\cite{Liao2appear}. At the leading order in LEFT, the short-distance contribution arises from effective interactions of dim-9 LNV operators that involve four quark fields and two charged lepton fields. To calculate the meson decay rate we match at the chiral symmetry breaking scale $\Lambda_\chi\approx 4\pi F_\pi$ the effective interactions of quarks to those of mesons that can be organized in chiral perturbation theory ($\chi$PT)~\cite{Gasser:1983yg,Gasser:1984gg}. The result thus obtained is general in the sense that it depends only on the Wilson coefficients of effective interactions in LEFT based on the QCD and QED gauge symmetries, and on the hadronic low energy constants (LECs) in $\chi$PT parameterizing nonperturbative QCD physics. These LECs may be extracted by chiral symmetry from other measured processes or computed in lattice theory. The merit in such an approach is that the uncertainties incurred in the result may be estimated systematically. This is in contrast to the studies in the literature~\cite{Abad:1984gh, Ivanov:2004ch,Cvetic:2010rw,  Abada:2017jjx,Li:2018pag, Chun:2019nwi,Quintero:2016iwi} where hadronic models or approximations such as vacuum insertion are appealed to estimate the hadronic matrix elements.

To translate the experimental constraints at low energy to those on new physics at a high scale, we have to climb up the ladder of energy scales in figure~\ref{fig1}. If there are no new particles with a mass at or below the electroweak scale $\Lambda_\textrm{EW}\approx m_W$, SM serves as a good starting point for an effective field theory, namely, the standard model effective field theory (SMEFT)~\cite{Weinberg:1979sa,
Buchmuller:1985jz,Grzadkowski:2010es,Lehman:2014jma,Liao:2016hru,
Liao:2019tep,Lehman:2015coa,Henning:2015alf,Henning:2017fpj}, between some new physics scale $\Lambda_\textrm{NP}$ and the scale $\Lambda_\textrm{EW}$. SMEFT enjoys the complete SM gauge symmetries $SU(3)_C\times SU(2)_L\times U(1)_Y$ but does not assume other symmetries such as lepton or baryon number conservation. When we match SMEFT and LEFT at $\Lambda_\textrm{EW}$ by integrating out heavy SM particles, i.e., the Higgs boson $h$, the weak gauge bosons $W^\pm,~Z$, and the top quark $t$, it turns out that the interaction structures entering the above general result for the decay rate simplify significantly. This simplification would disappear generically if one assumes a different EFT above the scale $\Lambda_\textrm{EW}$, such as the $\nu$SMEFT with relatively light sterile neutrinos~\cite{Aparici:2009fh,delAguila:2008ir,
Bhattacharya:2015vja,Liao:2016qyd} or the Higgs-Electroweak Chiral Lagrangian (EWCH$\mathcal{L}$)~\cite{Buchalla:2013rka}.

The outline of this paper is as follows. In section~\ref{sec2}, we first find out in LEFT a complete and independent basis for all dim-9 operators that violate lepton number $L$ by two units ($|\Delta L|=2$) relevant to the type of processes under consideration. Other seemingly independent operators are removed as redundant in appendix~\ref{app1}. Then, in section~\ref{sec3} we calculate the one-loop QCD contribution to the anomalous dimension matrix for the basis operators, and we correct as a byproduct the submatrix relevant to nuclear $0\nu\beta\beta$ decay obtained in the literature~\cite{Gonzalez:2015ady}. Our analytical and numerical solutions to the complete set of RGEs are delegated to appendix~\ref{app12}. In section~\ref{sec4} we match the above operators downwards the scale to those in $\chi$PT, and formulate a master formula for the decay width of $K^-\rightarrow\pi^+l^-l^-$. And in section~\ref{sec5} we do the opposite by matching the dim-9 operators in LEFT upwards the scale to dim-7 operators in SMEFT, and obtain the bounds on the relevant Wilson coefficients in SMEFT. Our conclusion is summarized in section~\ref{sec6}, while appendix~\ref{app2} and appendix~\ref{app3} reproduce respectively the dim-7 basis operators in SMEFT and renormalization group equations relevant to the decay under consideration.

%%%%%
\section{Basis for dim-9 LNV operators in LEFT}
\label{sec2}

Low energy effective field theory (LEFT) is an effective field theory for quarks and leptons that respects the QCD and QED gauge symmetries $SU(3)_C\times U(1)_{\text{EM}}$. If SM is considered as a fundamental theory or the leading interactions of an effective field theory (SMEFT), LEFT is the effective field theory below the electroweak scale $\Lambda_\textrm{EW}$ in which spontaneous symmetry breaking has already taken place and heavy particles like the Higgs and weak gauge bosons and the top quark have been integrated out. It has a natural expansion parameter of momentum (derivative) over $\Lambda_\textrm{EW}$, and the relative importance of effective interactions is judged by the canonical dimension of their operators.

For a consistent physics analysis it is important to establish first a basis of complete and independent operators. The short-distance contribution to the meson decays under consideration originates at leading order in LEFT from dim-9 LNV operators that involve four quark fields and two lepton fields. The basis of dim-9 operators relevant to nuclear $0\nu\beta\beta$ decay has been studied by several groups~\cite{Gonzalez:2015ady,Pas:2000vn,Prezeau:2003xn,Graesser:2016bpz}. For kaon decays, ref.~\cite{Quintero:2016iwi} suggested a set of operators by assigning a free flavor index to quark and lepton fields in the operators given in ref.~\cite{Gonzalez:2015ady}. As we will show below, this set is incomplete. We will make a thorough analysis on the basis of operators and verify our result by an independent approach based on the Hilbert series.

Since the operators with $L=+2$ are Hermitian conjugates of those with $L=-2$, it is sufficient to consider the latter which correspond to the decay $M_1^-\to M_2^+ l_\alpha^- l_\beta^-$ with $M_i^\pm$ being charged mesons. By Fierz identities, we can always reshuffle the fields to reach a quark-lepton separated form of operators:
\begin{eqnarray}
&&(\overline{u^p_{X_1}}\Gamma_1d^r_{Y_1})
[\overline{u^s_{X_2}}\Gamma_2d^t_{Y_2}]
(\overline{l_\alpha}\Gamma_3l_\beta^C),~~~
(\overline{u^p_{X_1}}\Gamma_1d^r_{Y_1}]
[\overline{u^s_{X_2}}\Gamma_2d^t_{Y_2})
(\overline{l_\alpha}\Gamma_3l_\beta^C).
\end{eqnarray}
Here $u^p_X$ ($d^p_X$) is an up-type (a down-type) quark field of flavor $p$ and chirality $X$, $l_\alpha$ a charged lepton field of flavor $\alpha$ and $l_\alpha^C$ is its charge conjugate. $\Gamma_i$ refers to the standard sixteen Dirac matrices, while the two brackets $(~,~)$ and $[~,~]$ indicate the two pairs of color contraction in the products of quark fields. We denote the lepton bilinears as
\begin{eqnarray}
&&j^{\alpha\beta}= (\overline{l_{\alpha}} l_{\beta}^C),~j_5^{\alpha\beta}= (\overline{l_{\alpha}}\gamma_5  l_{\beta}^C),~j_{5\mu}^{\alpha\beta}= (\overline{l_{\alpha}}\gamma_\mu\gamma_5  l_{\beta}^C),
\label{sylc}
\\
&&j_\mu^{\alpha\beta}=(\overline{l_{\alpha}}\gamma_\mu l_{\beta}^C),~j_{\mu\nu}^{\alpha\beta}= (\overline{l_{\alpha}}\sigma_{\mu\nu} l_{\beta}^C).
\label{antisylc}
\end{eqnarray}
The bilinears in equations~\eqref{sylc} and \eqref{antisylc} are respectively symmetric and antisymmetric under the interchange of the lepton flavors $\alpha$ and $\beta$. Thus the latter disappear for identical leptons but are generally present for different leptons. For simplicity we will focus on the decay into identical leptons $K^-\rightarrow\pi^+ l^-l^-$ in our later phenomenological analysis.

\begin{table}
\centering
\begin{tabular}{|l|l||l|l|}
\hline
Notation & Operator & Notation & Operator
\\
\hline
 $\calO_{prst}^{LLLL,S/P}$
 & $ (\overline{u_L^p}\gamma^\mu d_L^r)[\overline{u_L^s}\gamma_\mu d_L^t]( j^{\alpha\beta}/j_5^{\alpha\beta})$
 & $\calO_{prst}^{RRRR,S/P}$
 &  $(\overline{u_R^p}\gamma^\mu d_R^r)[\overline{u_R^s}\gamma_\mu d_R^t]( j^{\alpha\beta}/j_5^{\alpha\beta})$
 \\
 \hline
 $\calO_{prst}^{LLLL,T}$
 &  $ (\overline{u_L^p}\gamma^\mu d_L^r)[\overline{u_L^s}\gamma^\nu d_L^t]( j_{\mu\nu}^{\alpha\beta})$
 &  $\calO_{prst}^{RRRR,T}$
 &  $(\overline{u_R^p}\gamma^\mu d_R^r)[\overline{u_R^s}\gamma^\nu d_R^t]( j_{\mu\nu}^{\alpha\beta})$
  \\
 \hline
$\tilde{\calO}_{prst}^{LLLL,T}$
&  $(\overline{u_L^p}\gamma^\mu d_L^r][\overline{u_L^s}\gamma^\nu d_L^t)( j_{\mu\nu}^{\alpha\beta})$
&  $\tilde{\calO}_{prst}^{RRRR,T}$
&  $(\overline{u_R^p}\gamma^\mu d_R^r][\overline{u_R^s}\gamma^\nu d_R^t)( j_{\mu\nu}^{\alpha\beta})$
 \\
 \hline
$\calO_{prst}^{LRLR,S/P}$
& $(\overline{u_L^p}d_R^r)[\overline{u_L^s}d_R^t]
(j^{\alpha\beta}/j_5^{\alpha\beta})$
& $\calO_{prst}^{RLRL,S/P}$
& $(\overline{u_R^p}d_L^r)[\overline{u_R^s}d_L^t]
( j^{\alpha\beta}/j_5^{\alpha\beta})$
 \\
 \hline
$\tilde{\calO}_{prst}^{LRLR,S/P}$
& $(\overline{u_L^p}d_R^r][\overline{u_L^s}d_R^t)
( j^{\alpha\beta}/j_5^{\alpha\beta})$
& $\tilde{\calO}_{prst}^{RLRL,S/P}$
& $(\overline{u_R^p}d_L^r][\overline{u_R^s}d_L^t)
(j^{\alpha\beta}/j_5^{\alpha\beta}) $
 \\
 \hline
$\calO_{prst}^{LRLR,T}$
& $(\overline{u_L^p}i\sigma^{\mu\nu}d_R^r)[\overline{u_L^s}d_R^t] (j_{\mu\nu}^{\alpha\beta})$
& $\calO_{prst}^{RLRL,T}$
& $(\overline{u_R^p}i\sigma^{\mu\nu}d_L^r)[\overline{u_R^s}d_L^t]
(j_{\mu\nu}^{\alpha\beta})$
 \\
 \hline
$\tilde{\calO}_{prst}^{LRLR,T}$
& $(\overline{u_L^p}\sigma^{\mu\rho}d_R^r)
[\overline{u_L^s}\sigma^{\nu}_{~\rho}d_R^t]
(j_{\mu\nu}^{\alpha\beta})$
& $\tilde{\calO}_{prst}^{RLRL,T}$
& $(\overline{u_R^p}\sigma^{\mu\rho}d_L^r)
[\overline{u_R^s}\sigma^{\nu}_{~\rho}d_L^t]
(j_{\mu\nu}^{\alpha\beta})$
 \\
 \hline
 $\calO_{prst}^{LRLL,V/A}$
 & $(\overline{u_L^p}d_R^r)[\overline{u_L^s}\gamma^\mu d_L^t](j_\mu^{\alpha\beta} /j_{5\mu}^{\alpha\beta})$
 & $\calO_{prst}^{RLRR,V/A}$
 & $(\overline{u_R^p}d_L^r)[\overline{u_R^s}\gamma^\mu d_R^t]
 ( j_\mu^{\alpha\beta} /j_{5\mu}^{\alpha\beta})$
  \\
 \hline
 $\tilde{\calO}_{prst}^{LRLL,V/A}$
 & $(\overline{u_L^p}d_R^r][\overline{u_L^s}\gamma^\mu d_L^t)(j_\mu^{\alpha\beta} /j_{5\mu}^{\alpha\beta})$
 & $\tilde{\calO}_{prst}^{RLRR,V/A}$
 & $(\overline{u_R^p}d_L^r][\overline{u_R^s}\gamma^\mu d_R^t)(j_\mu^{\alpha\beta} /j_{5\mu}^{\alpha\beta})$
 \\
 \hline
$\calO_{prst}^{LRRR,V/A}$
& $(\overline{u_L^p}d_R^r)[\overline{u_R^s}\gamma^\mu d_R^t](j_\mu^{\alpha\beta} /j_{5\mu}^{\alpha\beta})$
& $\calO_{prst}^{RLLL,V/A}$
& $(\overline{u_R^p}d_L^r)[\overline{u_L^s}\gamma^\mu d_L^t]
( j_\mu^{\alpha\beta} /j_{5\mu}^{\alpha\beta})$
 \\
 \hline
$\tilde{\calO}_{prst}^{LRRR,V/A}$
& $(\overline{u_L^p}d_R^r][\overline{u_R^s}\gamma^\mu d_R^t)
(j_\mu^{\alpha\beta} /j_{5\mu}^{\alpha\beta})$
& $\tilde{\calO}_{prst}^{RLLL,V/A}$
& $(\overline{u_R^p}d_L^r][\overline{u_L^s}\gamma^\mu d_L^t)
(j_\mu^{\alpha\beta} /j_{5\mu}^{\alpha\beta})$
\\
\hline
$\calO_{prst}^{LRRL,T}$
& $(\overline{u_L^p}i\sigma^{\mu\nu}d_R^r)[\overline{u_R^s}d_L^t]
(j_{\mu\nu}^{\alpha\beta})$
& $\calO_{prst}^{RLLR,T}$
& $(\overline{u_R^p}i\sigma^{\mu\nu} d_L^r)[\overline{u_L^s}d_R^t](j_{\mu\nu}^{\alpha\beta})$
\\
\hline
$\tilde{\calO}_{prst}^{LRRL,T}$
& $(\overline{u_L^p}i\sigma^{\mu\nu}d_R^r][\overline{u_R^s}d_L^t)
(j_{\mu\nu}^{\alpha\beta})$
& $ \tilde{\calO}_{prst}^{RLLR,T}$
& $(\overline{u_R^p}i\sigma^{\mu\nu} d_L^r][\overline{u_L^s}d_R^t)(j_{\mu\nu}^{\alpha\beta})$
 \\
 \hline
 $\calO_{prst}^{LRRL,S/P}$
& $(\overline{u_L^p}d_R^r)[\overline{u_R^s}d_L^t]( j^{\alpha\beta}/j_5^{\alpha\beta})$
&&
\\
\hline
$\tilde{\calO}_{prst}^{LRRL,S/P}$
& $(\overline{u_L^p}d_R^r][\overline{u_R^s}d_L^t)
(j^{\alpha\beta}/j_5^{\alpha\beta})$
&&
\\
\hline
\end{tabular}
\caption{Dim-9 basis operators with $L=-2$. The chiralities of quarks and the Lorentz structure of the lepton bilinear are shown as superscripts of $\calO$ and the flavors of quarks as subscripts. $\tilde\calO$ differs from $\calO$ only in color contraction. A factor of $i$ is associated with $\sigma^{\mu\nu}$ so that the relevant anomalous dimension matrix elements in section~\ref{sec3} are real.}
\label{tab1}
\end{table}

By making judicious applications of the properties of Dirac matrices and Fierz identities derived in refs.~\cite{Liao:2016hru,Liao:2012uj}, we find all dim-9, $L=-2$ operators shown in table~\ref{tab1} according to the chirality ($L,~R$) of quark fields and the Lorentz structure ($S,~P,~V,~A,~T$) of lepton bilinears. The operators in the right column are parity partners of those in the left. There exist simple flavor symmetries in quark factors for some of the operators:
\begin{eqnarray}
&&\calO_{prst}^{LLLL,S/P}=\calO_{stpr}^{LLLL,S/P},~
\calO_{prst}^{LLLL,T}=-\calO_{stpr}^{LLLL,T},~
\tilde{\calO}_{prst}^{LLLL,T}=-\tilde{\calO}_{stpr}^{LLLL,T},
\\
&&\calO_{prst}^{RRRR,S/P}=\calO_{stpr}^{RRRR,S/P},~
\calO_{prst}^{RRRR,T}=-\calO_{stpr}^{RRRR,T},~
\tilde{\calO}_{prst}^{RRRR,T}=-\tilde{\calO}_{stpr}^{RRRR,T},
\\
&&\calO_{prst}^{LRLR,S/P}=\calO_{stpr}^{LRLR,S/P},~
\tilde{\calO}_{prst}^{LRLR,S/P}=\calO_{stpr}^{LRLR,S/P},~
\tilde{\calO}_{prst}^{LRLR,T}=-\tilde{\calO}_{stpr}^{LRLR,T},
\\
&&\calO_{prst}^{RLRL,S/P}=\calO_{stpr}^{RLRL,S/P},~
\tilde{\calO}_{prst}^{RLRL,S/P}=\calO_{stpr}^{RLRL,S/P},~
\tilde{\calO}_{prst}^{RLRL,T}=-\tilde{\calO}_{stpr}^{RLRL,T}.
\end{eqnarray}
Removing the redundant operators due to flavor symmetries, there are in total 5886 independent operators with five quarks and three charged leptons. A detailed discussion on removing redundancy is delegated to appendix~\ref{app1}, in which some apparently independent operators are shown to be actually redundant and decomposed into a linear combination of the basis operators in table~\ref{tab1}. We have verified our results by the Hilbert series approach in ref.~\cite{Henning:2015alf} which counts the number of independent operators for a specified set of fields but does not spell out their forms.

Let us compare our results with those in the literature. First, the operators with a tensor lepton bilinear are new to ref.~\cite{Quintero:2016iwi}, and moreover there are color exchanged operators that are not included in that reference. Second, if we restrict ourselves to the subspace of operators relevant to nuclear $0\nu\beta\beta$ decay, our results match the ones in refs.~\cite{Gonzalez:2015ady} and \cite{Graesser:2016bpz} when Fierz identities are applied. The relations among the three bases are
\begin{eqnarray}
&&\begin{pmatrix}
\calO_1^{LL}\\ \calO_2^{LL}
\end{pmatrix}
=\begin{pmatrix}
1 &  0\\
-4-\frac{8}{N} &  -8
\end{pmatrix}
\begin{pmatrix}
\calO_{2RL} \\ \calO_{2RL}^\lambda
\end{pmatrix}
=
\begin{pmatrix}
1 &  0\\
-4 &  -8
\end{pmatrix}
\begin{pmatrix}
\calO_{udud}^{RLRL,S/P} \\ \tilde{\calO}_{udud}^{RLRL,S/P}
\end{pmatrix},
\label{rel_in}
\\
%%%
&&\begin{pmatrix}
\calO_1^{RR}\\ \calO_2^{RR}
\end{pmatrix}
=\begin{pmatrix}
1 &  0\\
-4-\frac{8}{N} &  -8
\end{pmatrix}
\begin{pmatrix}
\calO_{2LR} \\ \calO_{2LR}^\lambda
\end{pmatrix}
=
\begin{pmatrix}
1 &  0\\
-4 &  -8
\end{pmatrix}
\begin{pmatrix}
\calO_{udud}^{LRLR,S/P} \\ \tilde{\calO}_{udud}^{LRLR,S/P}
\end{pmatrix},
\\
%%%
&&\begin{pmatrix}
\calO_1^{RL}\\\calO_3^{RL}
\end{pmatrix}
=\begin{pmatrix}
-\frac{1}{2N} &  -\frac{1}{2}\\
1 & 0
\end{pmatrix}
\begin{pmatrix}
\calO_{1LR} \\ \calO_{1LR}^\lambda
\end{pmatrix}
=\begin{pmatrix}
1 &  0\\
0 &  -2
\end{pmatrix}
\begin{pmatrix}
\calO_{udud}^{LRRL,S/P} \\ \tilde{\calO}_{udud}^{LRRL,S/P}
\end{pmatrix},
\\
%%%
&&\begin{pmatrix}
\calO_3^{LL}\\\calO_3^{RR}
\end{pmatrix}
=\begin{pmatrix}
\calO_{3L} \\ \calO_{3R}
\end{pmatrix}
=
\begin{pmatrix}
\calO_{udud}^{LLLL,S/P} \\ \calO_{udud}^{RRRR,S/P}
\end{pmatrix},
\\
%%%
&&\begin{pmatrix}
\calO_4^{LL}\\ \calO_5^{LL}
\end{pmatrix}
=\begin{pmatrix}
-\frac{N+2}{N} i &  -2i\\
1 &  0
\end{pmatrix}
\begin{pmatrix}
\calO_{LLRL}^\mu  \\ \calO_{LLRL}^{\lambda,\mu}
\end{pmatrix}
=\begin{pmatrix}
-i &  -2i\\
1 &  0
\end{pmatrix}
\begin{pmatrix}
\calO_{udud}^{RLLL,A} \\  \tilde{\calO}_{udud}^{RLLL,A}
\end{pmatrix},
\\
%%%
&&\begin{pmatrix}
\calO_4^{RR}\\\calO_5^{RR}
\end{pmatrix}
=\begin{pmatrix}
-\frac{N+2}{N} i &  -2i\\
1 &  0
\end{pmatrix}
\begin{pmatrix}
\calO_{RRLR}^\mu\\ \calO_{RRLR}^{\lambda,\mu}
\end{pmatrix}
=
\begin{pmatrix}
-i &  -2i\\
1 &  0
\end{pmatrix}
\begin{pmatrix}
\calO_{udud}^{LRRR,A}  \\ \tilde{\calO}_{udud}^{LRRR,A}
\end{pmatrix},
\\
%%%
&&\begin{pmatrix}
\calO_4^{LR} \\ \calO_5^{LR}
\end{pmatrix}
=\begin{pmatrix}
\frac{N+2}{N} i &  2i\\
1 &  0
\end{pmatrix}
\begin{pmatrix}
\calO_{LLLR}^\mu \\ \calO_{LLLR}^{\lambda,\mu}
\end{pmatrix}
=\begin{pmatrix}
i &  2i\\
1 &  0
\end{pmatrix}
\begin{pmatrix}
\calO_{udud}^{LRLL,A} \\  \tilde{\calO}_{udud}^{LRLL,A}
\end{pmatrix},
\\
%%%
&&\begin{pmatrix}
\calO_4^{RL} \\ \calO_5^{RL}
\end{pmatrix}
=\begin{pmatrix}
\frac{N+2}{N} i &  2i\\
1 &  0
\end{pmatrix}
\begin{pmatrix}
\calO_{RRRL}^\mu \\   \calO_{RRRL}^{\lambda,\mu}
\end{pmatrix}
=\begin{pmatrix}
i &  2i\\
1 &  0
\end{pmatrix}
\begin{pmatrix}
\calO_{udud}^{RLRR,A} \\  \tilde{\calO}_{udud}^{RLRR,A}
\end{pmatrix},
\label{rel_fi}
\end{eqnarray}
where the first two notations in each line refer respectively to refs.~\cite{Gonzalez:2015ady} and \cite{Graesser:2016bpz} and the last one is ours, and $N=3$ is color number. Since the operators in ref.~\cite{Graesser:2016bpz} do not include the lepton bilinear, we have implicitly striped the lepton bilinears from the other two bases in order to write down the above relations. We have also dropped a factor of 4 in the definition of operators in ref.~\cite{Gonzalez:2015ady}, which will not affect comparison of RGE results in section~\ref{sec3}.
These relations further confirm our results against ref.~\cite{Quintero:2016iwi}, and will be used in section~\ref{sec3} to clarify differences in one-loop QCD running to ref.~\cite{Gonzalez:2015ady}.

\begin{table}
\centering
\begin{tabular}{| l |c| l |c| l |c|}
\hline
~~~~~~~Decay & Exp. UL &~~~~~~~Decay & Exp. UL &~~~~~~~Decay & Exp. UL
\\
\hline
$K^-\rightarrow \pi^+ \mu^-\mu^-$
&$4.2\times10^{-11}$~\cite{CortinaGil:2019dnd}
&
$K^-\rightarrow \pi^+ e^-e^-$
&$2.2\times10^{-10}$~\cite{CortinaGil:2019dnd}
&
$K^-\rightarrow \pi^+ \mu^-e^-$
&$5.0\times10^{-10}$~\cite{Appel:2000tc}
\\
\hline
$D^-\rightarrow \pi^+ \mu^-\mu^-$
&$2.2\times10^{-8}$~\cite{Aaij:2013sua}
&
$D^-\rightarrow \pi^+ e^-e^-$
&$1.1\times10^{-6}$~\cite{Rubin:2010cq}
&
$D^-\rightarrow \pi^+ \mu^-e^-$
&$2.0\times10^{-6}$~\cite{Lees:2011hb}
\\
\hline
$D^-\rightarrow K^+ \mu^-\mu^-$
&$1.0\times10^{-5}$~\cite{Lees:2011hb}
&
$D^-\rightarrow K^+ e^-e^-$
&$9\times10^{-7}$~\cite{Lees:2011hb}
&
$D^-\rightarrow K^+ \mu^-e^-$
&$1.9\times10^{-6}$~\cite{Lees:2011hb}
\\
\hline
$D^-\rightarrow \rho^+ \mu^-\mu^-$
&$5.6\times10^{-4}$~\cite{Kodama:1995ia}
& $D^-\rightarrow \rho^+ e^-e^-$
& $-$
&$D^-\rightarrow \rho^+ \mu^-e^-$
& $-$
\\
\hline
$D^-\rightarrow K^{*+} \mu^-\mu^-$
& $8.5\times10^{-4}$~\cite{Kodama:1995ia}
& $D^-\rightarrow K^{*+} e^-e^-$
& $-$
& $D^-\rightarrow K^{*+} \mu^-e^-$
& $-$
\\
\hline
$D_s^-\rightarrow \pi^+ \mu^-\mu^-$
&$1.2\times10^{-7}$~\cite{Aaij:2013sua}
&
$D_s^-\rightarrow \pi^+ e^-e^-$
&$4.1\times10^{-6}$~\cite{Lees:2011hb}
&
$D_s^-\rightarrow \pi^+ \mu^-e^-$
&$8.4\times10^{-6}$~\cite{Lees:2011hb}
\\
\hline
$D_s^-\rightarrow K^+ \mu^-\mu^-$
&$1.3\times10^{-5}$~\cite{Lees:2011hb}
&
$D_s^-\rightarrow K^+ e^-e^-$
&$5.2\times10^{-6}$~\cite{Lees:2011hb}
&
$D_s^-\rightarrow K^+ \mu^-e^-$
&$6.1\times10^{-6}$~\cite{Lees:2011hb}
\\
\hline
$D_s^-\rightarrow K^{*+} \mu^-\mu^-$
&$1.4\times10^{-3}$~\cite{Kodama:1995ia}
& $D_s^-\rightarrow K^{*+} e^-e^-$
& $-$
&$D_s^-\rightarrow K^{*+} \mu^-e^-$
& $-$
\\
\hline
$B^-\rightarrow \pi^+ \mu^-\mu^-$
&$4.0\times10^{-9}$~\cite{Aaij:2014aba}
&
$B^-\rightarrow \pi^+ e^-e^-$
&$2.3\times10^{-8}$~\cite{BABAR:2012aa}
&
$B^-\rightarrow \pi^+ \mu^-e^-$
&$1.5\times10^{-7}$~\cite{Lees:2013gdj}
\\
\hline
$B^-\rightarrow K^+ \mu^-\mu^-$
&$4.1\times10^{-8}$~\cite{Aaij:2011ex}
&
$B^-\rightarrow K^+ e^-e^-$
&$3.0\times10^{-8}$~\cite{BABAR:2012aa}
&
$B^-\rightarrow K^+ \mu^-e^-$
&$1.6\times10^{-7}$~\cite{Lees:2013gdj}
\\
\hline
$B^-\rightarrow K^{*+} \mu^-\mu^-$
&$5.9\times10^{-7}$~\cite{Lees:2013gdj}
&
$B^-\rightarrow K^{*+} e^-e^-$
&$4.0\times10^{-7}$~\cite{Lees:2013gdj}
&
$B^-\rightarrow K^{*+} \mu^-e^-$
&$3.0\times10^{-7}$~\cite{Lees:2013gdj}
\\
\hline
$B^-\rightarrow \rho^{+} \mu^-\mu^-$
&$4.2\times10^{-7}$~\cite{Lees:2013gdj}
&
$B^-\rightarrow \rho^{+} e^-e^-$
&$1.7\times10^{-7}$~\cite{Lees:2013gdj}
&
$B^-\rightarrow \rho^{+} \mu^-e^-$
&$4.7\times10^{-7}$~\cite{Lees:2013gdj}
\\
\hline
$B^-\rightarrow D^+ \mu^-\mu^-$
&$6.9\times10^{-7}$~\cite{Aaij:2012zr}
&
$B^-\rightarrow D^+ e^-e^-$
&$2.6\times10^{-6}$~\cite{Seon:2011ni}
&
$B^-\rightarrow D^+ \mu^-e^-$
&$1.8\times10^{-6}$~\cite{Seon:2011ni}
\\
\hline
$B^-\rightarrow D_s^+ \mu^-\mu^-$
& $5.8\times10^{-7}$~\cite{Aaij:2012zr}
& $B^-\rightarrow  D_s^+ e^-e^-$
& $-$
&
$B^-\rightarrow D_s^+ \mu^-e^-$
& $-$
\\
\hline
$B^-\rightarrow D^{*+} \mu^-\mu^-$
&$2.4\times10^{-6}$~\cite{Aaij:2012zr}
&$B^-\rightarrow D^{*+} e^-e^-$
& $-$
&$B^-\rightarrow D^{*+} \mu^-e^-$
& $-$
\\
\hline
$\tau^-\rightarrow e^+\pi^-\pi^- $
&$2.0\times10^{-8}$~\cite{Miyazaki:2012mx}
&
$\tau^-\rightarrow e^+\pi^-K^-$
&$3.2\times10^{-8}$~\cite{Miyazaki:2012mx}
&
$\tau^-\rightarrow e^+K^- K^-$
&$3.3\times10^{-8}$~\cite{Miyazaki:2012mx}
\\
\hline
$\tau^-\rightarrow \mu^+ \pi^-\pi^-$
&$3.9\times10^{-8}$~\cite{Miyazaki:2012mx}
&
$\tau^-\rightarrow \mu^+ \pi^-K^-$
&$4.8\times10^{-8}$~\cite{Miyazaki:2012mx}
&
$\tau^-\rightarrow \mu^+ K^- K^-$
&$4.7\times10^{-8}$~\cite{Miyazaki:2012mx}
 \\
 \hline
\end{tabular}
\caption{Experimental upper limits (Exp. UL) on the LNV three-body decays of the mesons and $\tau$ lepton.}
\label{tab2}
\end{table}

Our basis of operators shown in table~\ref{tab1} is responsible for all leading-order short-distance mechanisms of low energy processes that violate lepton number by two units:
\begin{eqnarray}
\mathcal{L}_\textrm{LEFT}^{|\Delta L|=2}=
\sum_i C_i\calO_i+\textrm{H.C.},
\label{eff_in}
\end{eqnarray}
where the subscript $i$ covers all indices appearing in the operators. These processes include in particular the popular nuclear $0\nu\beta\beta$ decays and the three- and four-body decays of the charged mesons $K,~D,~D_s,~B$ and the $\tau$ lepton. The current experimental upper bounds on the three-body decays of the charged mesons and $\tau$ lepton are summarized in table~\ref{tab2}. Using the above effective Lagrangian we can match downwards the scale to heavy quark effective theory or $\chi$PT that is appropriate to the process under consideration, and the upper bounds then translate into constraints on the Wilson coefficients $C_i$. We can further match $\mathcal{L}_\textrm{LEFT}^{|\Delta L|=2}$ upwards the scale to an EFT such as SMEFT defined between the electroweak scale $\Lambda_\textrm{EW}$ and some new physics scale $\Lambda_\textrm{NP}$, so that we can set constraints on $\Lambda_\textrm{NP}$. In this manner all low energy data are connected through a sequence of EFTs to potential new physics defined at a high scale.

Before finishing this section we record here the 36 operators (involving 26 four-quark combinations) that contribute to the decay $K^-\rightarrow \pi^+l^-l^-$:
\begin{eqnarray}
\nonumber
&&\calO_{udus}^{LLLL,S/P},~\calO_{udus}^{LRLR,S/P},~
\tilde{\calO}_{udus}^{LRLR,S/P},~\calO_{uiuj}^{LRLL,A},~
\tilde{\calO}_{uiuj}^{LRLL,A},~\calO_{uiuj}^{LRRR,A},~
\tilde{\calO}_{uiuj}^{LRRR,A},~\calO_{udus}^{LRRL,S/P},~
\tilde{\calO}_{udus}^{LRRL,S/P},
\\
&&\calO_{udus}^{RRRR,S/P},~\calO_{udus}^{RLRL,S/P},~ \tilde{\calO}_{udus}^{RLRL,S/P},~\calO_{uiuj}^{RLRR,A},~
\tilde{\calO}_{uiuj}^{RLRR,A},~\calO_{uiuj}^{RLLL,A},~
\tilde{\calO}_{uiuj}^{RLLL,A},~\calO_{usud}^{LRRL,S/P},~
\tilde{\calO}_{usud}^{LRRL,S/P},
\label{ope_bas}
\end{eqnarray}
where $(i,j)=(d,s),~(s,d)$ and the operators in the second line are the parity partners in the first. For the decay $K^-\rightarrow\pi^+e^-\mu^-$ with different leptons in the final state, the antisymmetric lepton bilinears in equation~\eqref{antisylc} also enter. This will bring about a total of 70 dim-9 operators in LEFT, making their matching to $\chi$PT much more complicated. We thus defer its study to a separate work \cite{Liao2appear} in which we will also include the long-distance contribution to the decays.

%%%%%
\section{QCD RGEs for dim-9 LNV operators in LEFT}
\label{sec3}

\begin{figure}
\centering
\includegraphics[width=12cm]{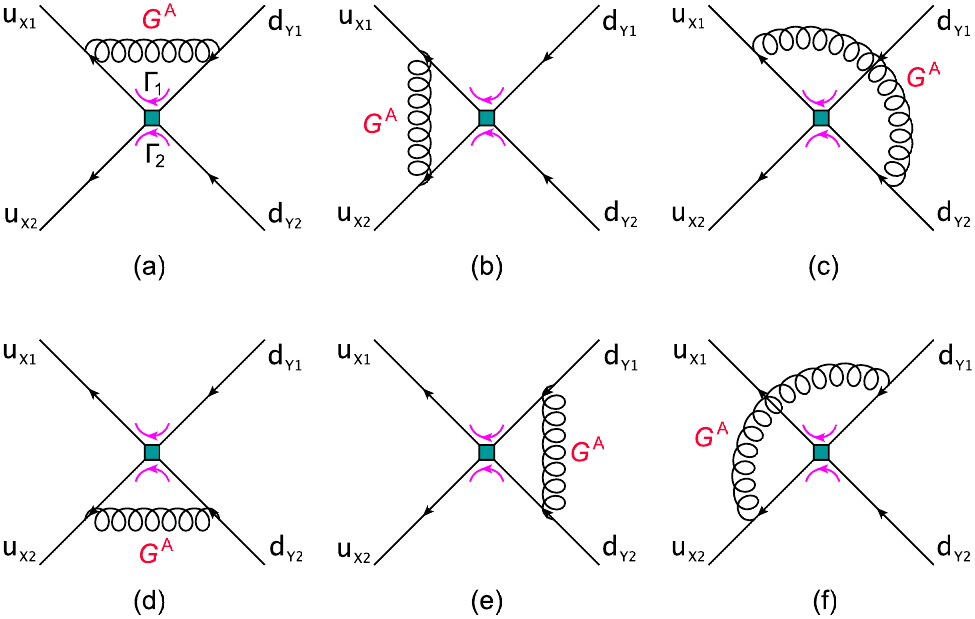}
\caption{One-loop QCD corrections to the four-quark operators $(\overline{u_{X_1}}\Gamma_1 d_{Y_1})(\overline{u_{X_2}}\Gamma_2 d_{Y_2})$.}
\label{fig2}
\end{figure}

To improve convergence in fixed-order perturbation theory we have to resum the large logarithms generated between two well-separated energy scales, i.e., $\Lambda_\textrm{EW}$ and $\Lambda_\chi$ for the processes under consideration. In this section we compute the one-loop QCD anomalous dimension matrix for all the LEFT operators shown in table~\ref{tab1} and equation~\eqref{eff_in}. The Feynman diagrams for the quark factors $(\overline{u_{X_1}}\Gamma_1 d_{Y_1})(\overline{u_{X_2}}\Gamma_2 d_{Y_2})$ of the basis operators are displayed in figure~\ref{fig2}. We perform the calculation with dimensional regularization and in the general $R_\xi$ gauge with gauge parameter $\xi_3$, and work with the $\overline{\text{MS}}$ scheme. The disappearance of $\xi_3$ in the final results will serve as a check of our calculation.

Since QCD conserves parity, it is sufficient to consider the operators listed in the left part of table~\ref{tab1}. Computing the Feynman diagrams in figure~\ref{fig2} and including the field wavefunction renormalization, we get the following one-loop QCD RGEs for the Wilson coefficients:
%%%%%
\begin{eqnarray}
\mu\frac{d}{d\mu}
\begin{pmatrix}
C_{prst}^{LLLL,S/P}
\\
C_{ptsr}^{LLLL,S/P}
\end{pmatrix}
&=&-\frac{\alpha_s}{2\pi}
\begin{pmatrix}
\frac{3}{N} & -3 \\
-3 &  \frac{3}{N}
\end{pmatrix}
\begin{pmatrix}
C_{prst}^{LLLL,S/P}
\\
C_{ptsr}^{LLLL,S/P}
\end{pmatrix},
\label{lrgi}
\\
\mu\frac{d}{d\mu}
\begin{pmatrix}
C_{prst}^{LLLL,T}
\\
\tilde{C}_{prst}^{LLLL,T}
\end{pmatrix}
&=&-\frac{\alpha_s}{2\pi}
\begin{pmatrix}
\frac{1}{N} & -1 \\
-1 &  \frac{1}{N}
\end{pmatrix}
\begin{pmatrix}
C_{prst}^{LLLL,T}
\\
\tilde{C}_{prst}^{LLLL,T}
\end{pmatrix},
\\
\mu\frac{d}{d\mu}
\begin{pmatrix}
C_{prst}^{LRLR,S/P}
\\
C_{ptsr}^{LRLR,S/P}
\\
\tilde{C}_{prst}^{LRLR,S/P}
\\
\tilde{C}_{ptsr}^{LRLR,S/P}
\end{pmatrix}
&=&-\frac{\alpha_s}{2\pi}
\begin{pmatrix}
\frac{2}{N}+6C_F & -4 & 2 & \frac{2}{N}-4C_F   \\
-4 & \frac{2}{N}+6C_F & \frac{2}{N}-4C_F &  2   \\
-2 & \frac{4}{N} & -\frac{2}{N}-2C_F & -2   \\
\frac{4}{N} & -2 & -2 & -\frac{2}{N}-2C_F
\end{pmatrix}
\begin{pmatrix}
C_{prst}^{LRLR,S/P}
\\
C_{ptsr}^{LRLR,S/P}
\\
\tilde{C}_{prst}^{LRLR,S/P}
\\
\tilde{C}_{ptsr}^{LRLR,S/P}
\end{pmatrix},
\\
\mu\frac{d}{d\mu}
\begin{pmatrix}
C_{prst}^{LRLR,T}
\\
C_{ptsr}^{LRLR,T}
\\
C_{srpt}^{LRLR,T}
\\
C_{stpr}^{LRLR,T}
\\
\tilde{C}_{prst}^{LRLR,T}
\end{pmatrix}
&=&-\frac{\alpha_s}{2\pi}
\begin{pmatrix}
2C_F & -1 & -1 & -\frac{2}{N}  & 0   \\
-1 & 2C_F &  -\frac{2}{N} & -1 & 0    \\
-1 &  -\frac{2}{N} & 2C_F & -1 & 0    \\
-\frac{2}{N} & -1 & -1 & 2C_F  & 0   \\
 0 & -\frac{1}{2}  & \frac{1}{2} & 0  & -2C_F    \\
\end{pmatrix}
\begin{pmatrix}
C_{prst}^{LRLR,T}
\\
C_{ptsr}^{LRLR,T}
\\
C_{srpt}^{LRLR,T}
\\
C_{stpr}^{LRLR,T}
\\
\tilde{C}_{prst}^{LRLR,T}
\end{pmatrix},
\\
\mu\frac{d}{d\mu}
\begin{pmatrix}
C_{prst}^{LRLL,V/A}
\\
C_{srpt}^{LRLL,V/A}
\\
\tilde{C}_{prst}^{LRLL,V/A}
\\
\tilde{C}_{srpt}^{LRLL,V/A}
\end{pmatrix}
&=&-\frac{\alpha_s}{2\pi}
\begin{pmatrix}
\frac{1}{N}+3C_F & -2 & 1 & \frac{1}{N}-2C_F   \\
-2 & \frac{1}{N}+3C_F & \frac{1}{N}-2C_F & 1  \\
-1 & \frac{2}{N} & -\frac{1}{N}-C_F & -1   \\
\frac{2}{N} & -1 & -1 & -\frac{1}{N}-C_F
\end{pmatrix}
\begin{pmatrix}
C_{prst}^{LRLL,V/A}
\\
C_{srpt}^{LRLL,V/A}
\\
\tilde{C}_{prst}^{LRLL,V/A}
\\
\tilde{C}_{srpt}^{LRLL,V/A}
\end{pmatrix},
\\
\mu\frac{d}{d\mu}
\begin{pmatrix}
C_{prst}^{LRRR,V/A}
\\
C_{ptsr}^{LRRR,V/A}
\\
\tilde{C}_{prst}^{LRRR,V/A}
\\
\tilde{C}_{ptsr}^{LRRR,V/A}
\end{pmatrix}
&=&-\frac{\alpha_s}{2\pi}
\begin{pmatrix}
\frac{1}{N}+3C_F & -2 & 1 & \frac{1}{N}-2C_F   \\
-2 & \frac{1}{N}+3C_F & \frac{1}{N}-2C_F & 1  \\
-1 & \frac{2}{N} & -\frac{1}{N}-C_F & -1   \\
\frac{2}{N} & -1 & -1 & -\frac{1}{N}-C_F
\end{pmatrix}
\begin{pmatrix}
C_{prst}^{LRRR,V/A}
\\
C_{ptsr}^{LRRR,V/A}
\\
\tilde{C}_{prst}^{LRRR,V/A}
\\
\tilde{C}_{ptsr}^{LRRR,V/A}
\end{pmatrix},
\\
\mu\frac{d}{d\mu}
\begin{pmatrix}
C_{prst}^{LRRL,T}
\\
\tilde{C}_{ptsr}^{LRRL,T}
\end{pmatrix}
&=&-\frac{\alpha_s}{2\pi}
\begin{pmatrix}
2C_F & 1 \\
0 &  -\frac{1}{N}
\end{pmatrix}
\begin{pmatrix}
C_{prst}^{LRRL,T}
\\
\tilde{C}_{ptsr}^{LRRL,T}
\end{pmatrix},
\\
\mu\frac{d}{d\mu}
\begin{pmatrix}
C_{prst}^{LRRL,S/P}
\\
\tilde{C}_{ptsr}^{LRRL,S/P}
\end{pmatrix}
&=&-\frac{\alpha_s}{2\pi}
\begin{pmatrix}
6C_F & 3 \\
0 &  -\frac{3}{N}
\end{pmatrix}
\begin{pmatrix}
C_{prst}^{LRRL,S/P}
\\
\tilde{C}_{ptsr}^{LRRL,S/P}
\end{pmatrix},
\label{lrgf}
\end{eqnarray}
where $\alpha_s=g_3^2/4\pi$ is the strong coupling constant and $C_F=(N^2-1)/2N=4/3$ the second Casimir invariant of quarks. We see that there remain no $\xi_3$-dependent terms. The general solutions to the above RGEs are given in appendix~\ref{app12}, together with numerical estimates running from the electroweak scale $\Lambda_\textrm{EW}$ to the chiral symmetry breaking scale $\Lambda_\chi$.

In ref.~\cite{Gonzalez:2015ady}, the one-loop QCD RGEs were computed for the subset of operators contributing to nuclear $0\nu\beta\beta$ decay. If we restrict ourselves to the same subset, our RGEs reduce to
%%%
\begin{eqnarray}
\mu\frac{d}{d\mu}C_{udud}^{LLLL,S/P}&=&
-\frac{\alpha_s}{2\pi}\Big(\frac{3}{N}-3 \Big)C_{udud}^{LLLL,S/P},
\\
\mu\frac{d}{d\mu}
\begin{pmatrix}
C_{udud}^{LRLR,S/P}
\label{run_in}
\\
\tilde{C}_{udud}^{LRLR,S/P}
\end{pmatrix}
&=&-\frac{\alpha_s}{2\pi}
\begin{pmatrix}
\frac{2}{N}+6C_F-4 & \frac{2}{N}-4C_F +2  \\
\frac{4}{N}-2 & -\frac{2}{N}-2C_F-2
\end{pmatrix}
\begin{pmatrix}
C_{udud}^{LRLR,S/P}
\\
\tilde{C}_{udud}^{LRLR,S/P}
\end{pmatrix},
\\
\mu\frac{d}{d\mu}
\begin{pmatrix}
C_{udud}^{LRLL,A}
\\
\tilde{C}_{udud}^{LRLL,A}
\end{pmatrix}
&=&-\frac{\alpha_s}{2\pi}
\begin{pmatrix}
\frac{1}{N}+3C_F-2 & \frac{1}{N}-2C_F+1   \\
\frac{2}{N}-1 & -\frac{1}{N}-C_F-1
\end{pmatrix}
\begin{pmatrix}
C_{udud}^{LRLL,A}
\\
\tilde{C}_{udud}^{LRLL,A}
\end{pmatrix},
\\
\mu\frac{d}{d\mu}
\begin{pmatrix}
C_{udud}^{LRRR,A}
\\
\tilde{C}_{udud}^{LRRR,A}
\end{pmatrix}
&=&-\frac{\alpha_s}{2\pi}
\begin{pmatrix}
\frac{1}{N}+3C_F-2 & \frac{1}{N}-2C_F+1   \\
\frac{2}{N}-1 &  -\frac{1}{N}-C_F-1
\end{pmatrix}
\begin{pmatrix}
C_{udud}^{LRRR,A}
\\
\tilde{C}_{udud}^{LRRR,A}
\end{pmatrix},
\\
\mu\frac{d}{d\mu}
\begin{pmatrix}
C_{udud}^{LRRL,S/P}
\\
\tilde{C}_{udud}^{LRRL,S/P}
\end{pmatrix}
&=&-\frac{\alpha_s}{2\pi}
\begin{pmatrix}
6C_F & 3 \\
0 &  -\frac{3}{N}
\end{pmatrix}
\begin{pmatrix}
C_{udud}^{LRRL,S/P}
\\
\tilde{C}_{udud}^{LRRL,S/P}
\end{pmatrix}.
\label{run_fi}
\end{eqnarray}
For comparison, we now recast our results in terms of the operator basis in ref.~\cite{Gonzalez:2015ady}. Consider two bases of operators $\calO_1$ and $\calO_2$ with Wilson coefficients $C_1$ and $C_2$ respectively, all of which are written in a column form. The algebraic equivalence, i.e., without appealing to integration by parts or equations of motion, of the bases implies $C_1^T\calO_{1}=C_2^T\calO_{2}$. Suppose the two bases are related by $\calO_1=V^T\calO_2$ where $V$ is nonsingular and contains pure numbers, we have $C_1=V^{-1}C_2$. The anomalous matrices computed in the two bases are then related by $\gamma_1=V^{-1}\gamma_2V$. For the case at hand, the matrix $V$ has been given in equations~\eqref{rel_in}-\eqref{rel_fi}. To summarize the result of comparison, we confirmed the anomalous dimension matrices $\hat{\gamma}^{XY}_{31}$, $\hat{\gamma}^{XY}_{12}$, and $\hat{\gamma}^{XX}_{3}$ in the notation of ref.~\cite{Gonzalez:2015ady}, but found differences for the other two matrices, for which we obtained
\begin{eqnarray}
\hat{\gamma}^{XX}_{45}
=-2\begin{pmatrix}
-\frac{3}{2}-C_F & -\frac{3i}{2}-\frac{3i}{N} \\
-\frac{i}{2}+\frac{i}{N} & 3C_F-\frac{3}{2} \\
\end{pmatrix},~
\hat{\gamma}^{XY}_{45}
=-2\begin{pmatrix}
-\frac{3}{2}-C_F & \frac{3i}{2}+\frac{3i}{N} \\
\frac{i}{2}-\frac{i}{N} & 3C_F-\frac{3}{2} \\
\end{pmatrix}.
\end{eqnarray}
Our $\hat{\gamma}^{XX}_{45}$ is half the result in ref.~\cite{Gonzalez:2015ady}, while $\hat{\gamma}^{XY}_{45}$ is completely different. We also computed those two matrices in the basis of that reference and confirmed our result.

%%%%%
\section{Matching onto chiral perturbation theory}
\label{sec4}

Our discussion on LEFT in the previous two sections applies generally to the case with five quarks (and all leptons). To study the specific hadronic process $K^-\rightarrow \pi^+l^-l^-$, we restrict ourselves to the LEFT with the three light quarks $u,~d,~s$ and match it at the scale $\Lambda_\chi$ to $\chi$PT for the Nambu-Goldstone bosons. The only guide we have for this matching calculation is the spontaneous chiral symmetry breaking $SU(3)_L\otimes SU(3)_R\to SU(3)_V$, whose consequences can be systematically worked out by $\chi$PT~\cite{Gasser:1983yg,Gasser:1984gg}; see refs.~\cite{Prezeau:2003xn,Graesser:2016bpz,Savage:1998yh,Cirigliano:2017ymo} for discussions in the case of lepton number violation. We will follow closely the technique clearly demonstrated in ref.~\cite{Graesser:2016bpz}.

In the matching to $\chi$PT the lepton bilinear of a dim-9 operator in LEFT behaves as a fixed external source, thus we only have to cope with the quark factor of the operator. Suppose the latter has been decomposed into a sum of irreducible representations (irreps) of the chiral group. A general irrep takes the form,
\begin{eqnarray}
\calO=T^{~ab}_{cd}(\overline{q_{X_1}^c}\Gamma_1 q_{Y_1,a})(\overline{q_{X_2}^d}\Gamma_2 q_{Y_2,b}),
\end{eqnarray}
where the set of pure numbers $T^{~ab}_{cd}$ depends on the irrep under consideration. $T$ is promoted as a spurion field that transforms properly together with chiral transformations of quarks,
\begin{eqnarray}
q_{L,a} \rightarrow L_a^{~p}q_{L,p},~
\overline{q_{R}}^b \rightarrow \overline{q_{R}}^p   (R^\dagger )_p^{~b}, ~q_{R,a} \rightarrow R_a^{~p}q_{R,p},~
\overline{q_{L}}^b \rightarrow \overline{q_{L}}^p   (L^\dagger )_p^{~b},
\end{eqnarray}
where $L\in SU(3)_L$ and $R\in SU(3)_R$, so that $\calO$ looks like a chiral invariant. On the $\chi$PT side, we introduce the standard matrix for the Nambu-Goldstone bosons (NGBs),
\begin{eqnarray}
\xi=\exp\left(\frac{i\Pi}{\sqrt{2}F_0}\right),~~~
\Pi=\begin{pmatrix}
\frac{\pi^0}{\sqrt{2}}+\frac{\eta}{\sqrt{6}} & \pi^+ & K^+
\\
\pi^- & -\frac{\pi^0}{\sqrt{2}}+\frac{\eta}{\sqrt{6}} & K^0
\\
K^- & \bar{K}^0 & -\sqrt{\frac{2}{3}}\eta
\end{pmatrix},
\end{eqnarray}
where $F_0$ is the decay constant in the chiral limit and $\xi^2\equiv\Sigma$. Under chiral transformations we have
\begin{eqnarray}
&&\xi_a^{~b} \rightarrow L_a^{~p}(\xi U^\dagger)_p^{~b} =(U\xi)_a^{~p} (R^\dagger)_p^{~b},~
\xi_a^{\dagger~b} \rightarrow R_a^{~p}(\xi^\dagger U^\dagger)_p^{~b}  =(U\xi^\dagger)_a^{~p} (L^\dagger)_p^{~b},
\end{eqnarray}
where $U\in SU(3)_V$ depends on the NGB fields. To form leading-order (LO) operators, matching is accomplished by the substitutions,
\begin{eqnarray}
q_{L,a}\rightarrow \xi_a^{~\alpha},~\overline{q_{L}}^a \rightarrow \xi_\alpha ^{\dagger~a},~
q_{R,a}\rightarrow \xi_a^{\dagger~\alpha}, ~\overline{q_{R}}^a \rightarrow \xi_\alpha^{~a},
\label{rep1}
\end{eqnarray}
where the free indices are to be contracted when forming an operator with $T^{~ab}_{cd}$. For hadronic operators appearing at the next-to-leading (NLO) or next-to-next-to-leading order (NNLO), covariant derivatives and quark masses will be involved in matching:
\begin{eqnarray}
&&q_{L,a}\rightarrow( (D_\mu\xi^\dagger)^\dagger )_a^{~\alpha},~
\overline{q_L}^a \rightarrow (D_\mu\xi^\dagger)_\alpha^{~a}, ~q_{R,a}\rightarrow (D_\mu\xi)_a^{\dagger \alpha},~\overline{q_{R}}^a \rightarrow (D_\mu\xi)_\alpha^{~a},
\\
&&q_{L,a}\rightarrow(M^\dagger \xi^\dagger)_a^{~\alpha},~
\overline{q_L}^a \rightarrow (\xi M)_\alpha^{~a}, ~
q_{R,a}\rightarrow ( M\xi)_a^{~\alpha},~
\overline{q_{R}}^a \rightarrow (\xi^\dagger M^\dagger)_\alpha^{~a},
\label{rep2}
\end{eqnarray}
where the quark mass $M={\diag}(m_u,~m_d,~m_s)$ as a spurion transforms like $M\to R^\dagger ML$ under the chiral group, and covariant derivatives transform as $D_\mu\xi\rightarrow U D_\mu\xi R^\dagger,~D_\mu\xi^\dagger \rightarrow UD_\mu\xi^\dagger L^\dagger$, with $D_\mu=\partial_\mu+(\xi^\dagger \partial_\mu \xi+ \xi\partial_\mu\xi^\dagger )/2$. They are related to the ordinary derivative of the $\Sigma$ field by $\xi D_\mu\xi^\dagger=(\Sigma\partial_\mu \Sigma^\dagger)/2$ and $\xi^\dagger(D_\mu\xi)=(\Sigma^\dagger\partial_\mu\Sigma)/2$. The identities $(D_\mu \xi)\xi^\dagger=-\xi (D_\mu \xi)^\dagger$ and $\xi^\dagger(D_\mu \xi)
=-(D_\mu \xi)^\dagger\xi$ are useful when reducing redundant operators.
\footnote{There is a disagreement here with ref.~\cite{Graesser:2016bpz}: we have a single mesonic operator for the irrep ${\bf 27}_L\times {\bf 1}_R$ as shown in table~\ref{tab3}, while that reference has six operators for the corresponding $\calO_{3L}$ consisting of three single-trace ones and three double-trace ones (equations~[4.6] and [4.7]). Our result is consistent with the other three papers: equations~(8) and (9) in ref.~\cite{Savage:1998yh}, equation~(33) in ref.~\cite{Prezeau:2003xn} (for the parity-even operator, i.e., ${\bf 27}_L\times {\bf 1}_R+{\bf 1}_L\times{\bf 27}_R$), and equation~(15c) in ref.~\cite{Cirigliano:2017ymo}. The problem seems to arise from the misidentification of chiral transformations of $\xi$ with $D_\mu\xi$ in ref.~\cite{Graesser:2016bpz}: while $\xi\to L\xi U^\dagger=U\xi R^\dagger$, $D_\mu\xi$ transforms only as $D_\mu\xi\to U(D_\mu\xi)R^\dagger$.}

\begin{table}[!ht]
\centering
\begin{tabular}{|c|c|c|c|}
\hline
Notation
& Quark operator
& chiral  irrep
& Hadronic operator
\\
\hline
 $\calO_{udus}^{LLLL,S/P}~(\checkmark)$
 & $(\overline{u_L}\gamma^\mu d_L)[\overline{u_L}\gamma_\mu s_L](j/j_5)$
 & ${\bf 27}_L\times {\bf1}_R$
 & $\frac{5}{12}g_{27\times 1}F_0^4(\Sigma i\partial_\mu \Sigma^\dagger)_2^{~1}(\Sigma i\partial^\mu \Sigma^\dagger)_3^{~1}$
 \\
 \hline
 $\calO_{udus}^{RRRR,S/P}~(P)$
 & $(\overline{u_R}\gamma^\mu d_R)[\overline{u_R}\gamma_\mu s_R](j/j_5)$
 & ${\bf1}_L \times {\bf 27}_R$
  & $\frac{5}{12}g_{1\times 27}F_0^4(\Sigma^\dagger i\partial_\mu \Sigma)_2^{~1}(\Sigma^\dagger i\partial^\mu \Sigma)_3^{~1}$
 \\
 \hline
 $\calO_{udus}^{LRLR,S/P}~(\checkmark)$
 &  $(\overline{u_L} d_R)[\overline{u_L}s_R](j/j_5)$
 &  $ \overline{\bf 6}_L\times {\bf 6}_R$
 & $-g_{\overline{6}\times6}^a\frac{F_0^4}{4}(\Sigma^\dagger)_2^{~1}
 (\Sigma^\dagger)_3^{~1}$
  \\
 \hline
$ \tilde{\calO}_{udus}^{LRLR,S/P}~(\checkmark)$
 &  $(\overline{u_L}d_R][\overline{u_L}s_R)(j/j_5)$
 &  $\overline{\bf 6}_L\times {\bf 6}_R$
 & $-g_{\overline{6}\times6}^b\frac{F_0^4}{4}(\Sigma^\dagger)_2^{~1}
 (\Sigma^\dagger)_3^{~1}$
 \\
 \hline
  $\calO_{udus}^{RLRL,S/P}~(P)$
 &  $(\overline{u_R} d_L)[\overline{u_R}s_L](j/j_5)$
 &  $ {\bf 6}_L\times\overline{\bf 6}_R$
  & $-g_{6\times\overline{6}}^a\frac{F_0^4}{4}(\Sigma)_2^{~1}
  (\Sigma)_3^{~1}$
  \\
 \hline
$ \tilde{\calO}_{udus}^{RLRL,S/P}~(P)$
 &  $(\overline{u_R}d_L][\overline{u_R}s_L)(j/j_5)$
 &  ${\bf 6}_L\times\overline{\bf 6}_R $
   & $-g_{6\times\overline{6}}^b\frac{F_0^4}{4}(\Sigma)_2^{~1}
   (\Sigma)_3^{~1}$
 \\
 \hline
$ \calO_{udus}^{LRLL,A}~(\checkmark)$
& $(\overline{u_L}d_R)[\overline{u_L}\gamma^\mu s_L]j_{\mu 5} $
& $\overline{\bf 15}_L\times {\bf 3}_R$
& $-g_{\overline{15}\times3}^a\frac{F_0^4}{4}(\Sigma i\partial_\mu \Sigma^\dagger)_3^{~1}(\Sigma^\dagger)_2^{~1}$
 \\
 \hline
$\tilde{\calO}_{udus}^{LRLL,A}~(\checkmark)$
& $(\overline{u_L} d_R][\overline{u_L}\gamma^\mu s_L)j_{\mu 5}$
& $ \overline{\bf 15}_L\times {\bf 3}_R$
& $-g_{\overline{15}\times3}^b\frac{F_0^4}{4}(\Sigma i\partial_\mu \Sigma^\dagger)_3^{~1}(\Sigma^\dagger)_2^{~1}$
 \\
 \hline
 $ \calO_{usud}^{LRLL,A}~(\checkmark)$
& $(\overline{u_L}s_R)[\overline{u_L}\gamma^\mu d_L]j_{\mu 5} $
& $\overline{\bf 15}_L\times {\bf 3}_R$
& $-g_{\overline{15}\times3}^c\frac{F_0^4}{4}(\Sigma i\partial_\mu \Sigma^\dagger)_2^{~1}(\Sigma^\dagger)_3^{~1}$
 \\
 \hline
$\tilde{\calO}_{usud}^{LRLL,A}~(\checkmark)$
& $(\overline{u_L} s_R][\overline{u_L}\gamma^\mu d_L)j_{\mu 5}$
& $\overline{\bf 15}_L\times {\bf 3}_R$
& $-g_{\overline{15}\times3}^d\frac{F_0^4}{4}(\Sigma i\partial_\mu \Sigma^\dagger)_2^{~1}(\Sigma^\dagger)_3^{~1}$
 \\
 \hline
 $ \calO_{udus}^{RLRR,A}~(P)$
& $(\overline{u_R}d_L)[\overline{u_R}\gamma^\mu s_R]j_{\mu 5} $
& ${\bf 3}_L\times\overline{\bf 15}_R$
& $-g_{3\times\overline{15}}^a\frac{F_0^4}{4}(\Sigma^\dagger i\partial_\mu \Sigma)_3^{~1}(\Sigma)_2^{~1}$
 \\
 \hline
$\tilde{\calO}_{udus}^{RLRR,A}~(P)$
& $(\overline{u_R} d_L][\overline{u_R}\gamma^\mu s_R)j_{\mu 5}$
& ${\bf 3}_L\times\overline{\bf 15}_R$
& $-g_{3\times\overline{15}}^b\frac{F_0^4}{4}(\Sigma^\dagger i\partial_\mu \Sigma)_3^{~1}(\Sigma)_2^{~1}$
 \\
 \hline
  $ \calO_{usud}^{RLRR,A}~(P)$
& $(\overline{u_R}s_L)[\overline{u_R}\gamma^\mu d_R]j_{\mu 5} $
& ${\bf 3}_L\times\overline{\bf 15}_R$
& $-g_{3\times\overline{15}}^c\frac{F_0^4}{4}(\Sigma^\dagger i\partial_\mu \Sigma)_2^{~1}(\Sigma)_3^{~1}$
 \\
 \hline
$\tilde{\calO}_{usud}^{RLRR,A}~(P)$
& $(\overline{u_R} s_L][\overline{u_R}\gamma^\mu d_R)j_{\mu 5}$
& $ {\bf 3}_L\times\overline{\bf 15}_R$
& $-g_{3\times\overline{15}}^d\frac{F_0^4}{4}(\Sigma^\dagger i\partial_\mu \Sigma)_2^{~1}(\Sigma)_3^{~1}$
 \\
 \hline
$\calO_{udus+}^{LRRR,A}~(\checkmark)$
& $\frac{1}{2}\Big[ (\overline{u_L}d_R)[\overline{u_R}\gamma^\mu s_R]+d\leftrightarrow s \Big]j_{\mu 5}$
& $\overline{\bf 3}_L\times{\bf 15}_R$
& $g_{\overline{3}\times15}^a\frac{F_0^4}{4}\Big[(\Sigma^\dagger)_2^{~1}
(\Sigma^\dagger i\partial_\mu \Sigma)_3^{~1}
+(\Sigma^\dagger)_3^{~1}(\Sigma^\dagger i\partial_\mu \Sigma)_2^{~1} \Big]$
 \\
 \hline
  $\tilde{\calO}_{udus+}^{LRRR,A}~(\checkmark)$
 & $\frac{1}{2}\Big[ (\overline{u_L}d_R][\overline{u_R}\gamma^\mu s_R)+d\leftrightarrow s \Big]j_{\mu 5}$
 & $\overline{\bf 3}_L\times{\bf 15}_R$
 & $g_{\overline{3}\times15}^b\frac{F_0^4}{4}\Big[(\Sigma^\dagger)_2^{~1}
 (\Sigma^\dagger i\partial_\mu \Sigma)_3^{~1}
 +(\Sigma^\dagger)_3^{~1}(\Sigma^\dagger i\partial_\mu \Sigma)_2^{~1} \Big]$
  \\
 \hline
$\calO_{udus-}^{LRRR,A}~(\checkmark)$
& $\frac{1}{2}\Big[ (\overline{u_L}d_R)[\overline{u_R}\gamma^\mu s_R]-d\leftrightarrow s \Big]j_{\mu 5}$
& $\overline{\bf 3}_L\times\overline{\bf 6}_R$
& $g_{\overline{3}\times\overline{6}}^a\frac{F_0^4}{4}
\Big[(\Sigma^\dagger)_2^{~1}(\Sigma^\dagger i\partial_\mu \Sigma)_3^{~1}-(\Sigma^\dagger)_3^{~1}(\Sigma^\dagger i\partial_\mu \Sigma)_2^{~1} \Big]$
 \\
 \hline
 $\tilde{\calO}_{udus-}^{LRRR,A}~(\checkmark)$
 & $\frac{1}{2}\Big[ (\overline{u_L}d_R][\overline{u_R}\gamma^\mu s_R)-d\leftrightarrow s\Big]j_{\mu 5}$
 & $\overline{\bf 3}_L\times\overline{\bf 6}_R$
 & $g_{\overline{3}\times\overline{6}}^b\frac{F_0^4}{4}
 \Big[(\Sigma^\dagger)_2^{~1}(\Sigma^\dagger i\partial_\mu \Sigma)_3^{~1}-(\Sigma^\dagger)_3^{~1}(\Sigma^\dagger i\partial_\mu \Sigma)_2^{~1} \Big]$
 \\
 \hline
 $\calO_{udus+}^{RLLL,A}~(P)$
& $\frac{1}{2}\Big[ (\overline{u_R}d_L)[\overline{u_L}\gamma^\mu s_L]+d\leftrightarrow s \Big]j_{\mu 5}$
& ${\bf 15}_L\times\overline{\bf 3}_R$
 & $g_{15\times\overline{3}}^a\frac{F_0^4}{4}\Big[(\Sigma)_2^{~1}(\Sigma i\partial_\mu \Sigma^\dagger)_3^{~1}+(\Sigma)_3^{~1}(\Sigma i\partial_\mu \Sigma^\dagger)_2^{~1} \Big]$
 \\
 \hline
  $\tilde{\calO}_{udus+}^{RLLL,A}~(P)$
 & $\frac{1}{2}\Big[ (\overline{u_R}d_L][\overline{u_L}\gamma^\mu s_L)+d\leftrightarrow s \Big]j_{\mu 5}$
 & ${\bf 15}_L\times\overline{\bf 3}_R$
  & $g_{15\times\overline{3}}^b\frac{F_0^4}{4}\Big[(\Sigma)_2^{~1}(\Sigma i\partial_\mu \Sigma^\dagger)_3^{~1}+(\Sigma)_3^{~1}(\Sigma i\partial_\mu \Sigma^\dagger)_2^{~1} \Big]$
  \\
 \hline
$\calO_{udus-}^{RLLL,A}~(P)$
& $\frac{1}{2}\Big[ (\overline{u_R}d_L)[\overline{u_L}\gamma^\mu s_L]-d\leftrightarrow s \Big]j_{\mu 5}$
& $\overline{\bf 6}_L\times\overline{\bf 3}_R$
& $g_{\overline{6}\times\overline{3}}^a\frac{F_0^4}{4}
\Big[(\Sigma)_2^{~1}(\Sigma i\partial_\mu \Sigma^\dagger)_3^{~1}-(\Sigma)_3^{~1}(\Sigma i\partial_\mu \Sigma^\dagger)_2^{~1} \Big]$
 \\
 \hline
 $\tilde{\calO}_{udus-}^{RLLL,A}~(P)$
 & $\frac{1}{2}\Big[ (\overline{u_R}d_L][\overline{u_L}\gamma^\mu s_L)-d\leftrightarrow s\Big]j_{\mu 5}$
 & $\overline{\bf 6}_L\times\overline{\bf 3}_R$
 & $g_{\overline{6}\times\overline{3}}^b\frac{F_0^4}{4}
 \Big[(\Sigma)_2^{~1}(\Sigma i\partial_\mu \Sigma^\dagger)_3^{~1}-(\Sigma)_3^{~1}(\Sigma i\partial_\mu \Sigma^\dagger)_2^{~1} \Big]$
 \\
 \hline
$\calO_{udus}^{LRRL,S/P}~(\checkmark)$
& $(\overline{u_L} d_R)[\overline{u_R}s_L](j/j_5)$
& ${\bf 8}_L\times {\bf 8}_R$
& $g_{8\times 8}^a\frac{F_0^4}{4} (\Sigma^\dagger)_2^{~1}(\Sigma)_3^{~1}$
 \\
 \hline
$\tilde{\calO}_{udus}^{LRRL,S/P}~(\checkmark)$
& $(\overline{u_L} d_R][\overline{u_R}s_L)(j/j_5)$
& ${\bf 8}_L\times {\bf 8}_R$
& $g_{8\times 8}^b\frac{F_0^4}{4} (\Sigma^\dagger)_2^{~1}(\Sigma)_3^{~1}$
\\
\hline
$\calO_{usud}^{LRRL,S/P}~(P)$
& $(\overline{u_L} s_R)[\overline{u_R}d_L](j/j_5)$
& ${\bf 8}_L\times {\bf 8}_R$
& $g_{8\times 8}^c\frac{F_0^4}{4} (\Sigma^\dagger)_3^{~1}(\Sigma)_2^{~1}$
 \\
 \hline
$\tilde{\calO}_{usud}^{LRRL,S/P}~(P)$
& $(\overline{u_L} s_R][\overline{u_R}d_L)(j/j_5)$
& ${\bf 8}_L\times {\bf 8}_R$
& $g_{8\times 8}^d\frac{F_0^4}{4} (\Sigma^\dagger)_3^{~1}(\Sigma)_2^{~1}$
\\
\hline
\end{tabular}
\caption{Quark factors of dim-9 operators in LEFT are matched at the leading nonvanishing order to their hadronic counterparts in $\chi$PT that are relevant to $K^-\rightarrow\pi^+l^-l^-$. Nonvanishing lepton bilinears $j=(\overline{l}l^C),~j_5= (\overline{l}\gamma_5 l^C),~j_{\mu 5}= (\overline{l}\gamma_\mu\gamma_5l^C)$ for an identical lepton pair $l=e,~\mu$ are not shown in hadronic factors for brevity.}
\label{tab3}
\end{table}

Now we turn to the decay $K^-\rightarrow\pi^+l^-l^-$. According to the procedure outlined above, we first decompose the quark factors in operators of equation~\eqref{ope_bas} into irreps under the chiral group, and then we obtain their leading nonvanishing hadronic counterparts according to equations~\eqref{rep1}-\eqref{rep2}. Our results are shown in table~\ref{tab3}. The matching coefficients are denoted by the LECs, $g_X$. These constants are difficult to compute because of strong dynamics, but some of them may be related by chiral symmetry to other constants that have been experimentally measured or computed by lattice simulations. In particular, the matrix elements of $K^-\rightarrow \pi^+$ associated with $g_{27\times 1}$, $g_{\overline 6 \times 6}^i$, and $g_{8\times 8}^i$ can be related to those of $\pi^-\rightarrow \pi^+$ which are responsible for the short-range mechanism of nuclear $0\nu\beta\beta$ decay and computed in \cite{Cirigliano:2017ymo}, and to those of $K^+\rightarrow\pi^+\pi^0$~\cite{Savage:1998yh} and $K^0\rightarrow\bar{K}^0$~\cite{Cirigliano:2017ymo} which have been computed by lattice simulations in~\cite{Carrasco:2015pra,Bertone:2012cu,
Boyle:2012qb,Jang:2015sla,Garron:2016mva} and~\cite{Blum:2012uk,Blum:2015ywa}, respectively. Here we adopt the results from~\cite{Cirigliano:2017djv}; in our notations:
\begin{eqnarray}
g_{27\times 1}=0.38\pm 0.08,~
g_{8\times 8}^a=5.5\pm2~{\GeV}^2,~
g_{8\times 8}^b=1.55\pm0.65~{\GeV}^2.
\label{lecs}
\end{eqnarray}
We make some clarifications concerning the results in table~\ref{tab3}. An operator with a symbol $(P)$ is the parity partner of a corresponding operator with a $(\checkmark)$. Parity invariance of QCD implies that such a pair of operators shares the same LEC:
\begin{eqnarray}
\nonumber
&&
g_{1\times 27}=g_{27\times 1},~g^a_{6\times \overline{6}}=g^a_{\overline{6}\times 6},~
g^b_{6\times \overline{6}}=g^b_{\overline{6}\times 6},~g^a_{3\times \overline{15}}=g^a_{\overline{15}\times 3},
\\
&&
g^b_{3\times \overline{15}}=g^b_{\overline{15}\times 3},~g^c_{3\times \overline{15}}=g^c_{\overline{15}\times 3},~
g^d_{3\times \overline{15}}=g^d_{\overline{15}\times 3},~g^a_{15\times \overline{3}}=g^a_{\overline{3}\times 15},
\\
\nonumber
&&
g^b_{15\times \overline{3}}=g^b_{\overline{3}\times 15},~g^a_{ \overline{6}\times \overline{3}}=g^a_{\overline{3}\times \overline{6}},~
g^b_{ \overline{6}\times \overline{3}}=g^b_{\overline{3}\times \overline{6}},~g^c_{8\times8}=g^a_{8\times8},
~g^d_{8\times8}=g^b_{8\times8}.
\end{eqnarray}
This leaves us with 13 LECs. The operators $\calO_{udus}^{LRLL,A}$ and $\calO_{usud}^{LRLL,A}$ have the same types of color contraction and belong to the same irrep of the chiral group but have the $d,~s$ quarks interchanged. They should have the same LEC upon ignoring the mass difference of the $d,~s$ quarks, $g^c_{\overline{15}\times 3}\approx g^a_{\overline{15}\times 3}$; similarly, we have $g^d_{\overline{15}\times 3}\approx g^b_{\overline{15}\times 3}$. Note that operators in the same chiral irrep but with different types of color contraction have generally different LECs. We thus have generally 11 LECs under a reasonable approximation.

We are now ready to write down the effective Lagrangian contributing to the decay $K^-\rightarrow \pi^+l^-l^-$ according to equations~\eqref{eff_in}-\eqref{ope_bas} and table~\ref{tab3} at the leading order of each operator:
\begin{eqnarray}
\nonumber
\mathcal{L}_{K^-\rightarrow \pi^+l^-l^-}&=&
 \frac{1}{2}K^-\pi^-\left[ c_1\left(\bar{l}l^C\right)+c_2\left(\bar{l}\gamma_5l^C\right)\right]
+\frac{1}{2}\left[ c_3\partial^\mu K^-\pi^- +c_4\partial^\mu \pi^-K^-\right]\left(\bar{l}\gamma_\mu \gamma_5l^C\right)
\\
&&
+\frac{1}{2}\partial^\mu K^-\partial_\mu\pi^-\left[ c_5\left(\bar{l}l^C\right)+c_6\left(\bar{l}\gamma_5l^C\right)\right],
\label{eff_in2}
\end{eqnarray}
where the parameters $c_i$ are
\begin{eqnarray}
\nonumber
c_1&=&g_{\overline{6}\times6}^aF_0^2
\left(C_{udus}^{LRLR,S}+C_{udus}^{RLRL,S} \right)
+g_{\overline{6}\times6}^bF_0^2
\left(\tilde{C}_{udus}^{LRLR,S}+\tilde{C}_{udus}^{RLRL,S} \right)
\label{c1}
\\
&&+g_{8\times8}^aF_0^2\left(C_{udus}^{LRRL,S}+C_{usud}^{LRRL,S} \right)
+g_{8\times8}^bF_0^2\left(\tilde{C}_{udus}^{LRRL,S}
+\tilde{C}_{usud}^{LRRL,S} \right),
\\\nonumber
c_2&=&g_{\overline{6}\times6}^aF_0^2\left(C_{udus}^{LRLR,P}
+C_{udus}^{RLRL,P} \right)
+g_{\overline{6}\times6}^bF_0^2\left(\tilde{C}_{udus}^{LRLR,P}
+\tilde{C}_{udus}^{RLRL,P} \right)
\label{c2}
\\
&&+g_{8\times8}^aF_0^2\left(C_{udus}^{LRRL,P}+C_{usud}^{LRRL,P} \right)
+g_{8\times8}^bF_0^2\left(\tilde{C}_{udus}^{LRRL,P}
+\tilde{C}_{usud}^{LRRL,P} \right),
\\\nonumber %%
c_3&=&
i g_{\overline{15}\times3}^aF_0^2
\left(C_{usud}^{LRLL,A}+C_{usud}^{RLRR,A}\right)
+i g_{\overline{15}\times3}^b F_0^2\left(\tilde{C}_{usud}^{LRLL,A}+\tilde{C}_{usud}^{RLRR,A} \right)
\\
\nonumber
&&+ig_{\overline{3}\times15}^aF_0^2\left(C_{udus}^{LRRR,A}
+C_{usud}^{LRRR,A}+C_{udus}^{RLLL,A}+C_{usud}^{RLLL,A}\right)
\\
\nonumber %%
&&+ig_{\overline{3}\times15}^bF_0^2\left(\tilde{C}_{udus}^{LRRR,A}
+\tilde{C}_{usud}^{LRRR,A}
+\tilde{C}_{udus}^{RLLL,A}+\tilde{C}_{usud}^{RLLL,A} \right)
\\
\nonumber
&&+ig_{\overline{3}\times\overline{6}}^aF_0^2
\left(C_{udus}^{LRRR,A}-C_{usud}^{LRRR,A}
+C_{udus}^{RLLL,A}-C_{usud}^{RLLL,A} \right)
\\
&&+ig_{\overline{3}\times\overline{6}}^bF_0^2
\left(\tilde{C}_{udus}^{LRRR,A}-\tilde{C}_{usud}^{LRRR,A}
+\tilde{C}_{udus}^{RLLL,A}-\tilde{C}_{usud}^{RLLL,A} \right),
\label{c3}
\\
\nonumber
c_4&=&
i g_{\overline{15}\times3}^cF_0^2
\left(C_{udus}^{LRLL,A}+C_{udus}^{RLRR,A} \right)
+i g_{\overline{15}\times3}^dF_0^2
\left(\tilde{C}_{udus}^{LRLL,A}+\tilde{C}_{udus}^{RLRR,A} \right)
\\
\nonumber
&&+ig_{\overline{3}\times15}^aF_0^2
\left(C_{udus}^{LRRR,A}+C_{usud}^{LRRR,A}
+C_{udus}^{RLLL,A}+C_{usud}^{RLLL,A} \right)
\\
\nonumber
&&+ig_{\overline{3}\times15}^bF_0^2\left(\tilde{C}_{udus}^{LRRR,A}
+\tilde{C}_{usud}^{LRRR,A}
+\tilde{C}_{udus}^{RLLL,A}+\tilde{C}_{usud}^{RLLL,A} \right)
\\
\nonumber
&&-ig_{\overline{3}\times\overline{6}}^aF_0^2
\left(C_{udus}^{LRRR,A}-C_{usud}^{LRRR,A}
+C_{udus}^{RLLL,A}-C_{usud}^{RLLL,A} \right)
\\
&&-ig_{\overline{3}\times\overline{6}}^bF_0^2
\left(\tilde{C}_{udus}^{LRRR,A}-\tilde{C}_{usud}^{LRRR,A}
+\tilde{C}_{udus}^{RLLL,A}-\tilde{C}_{usud}^{RLLL,A} \right),
\label{c4}
\\ %%
c_5&=&\frac{5}{3} g_{27\times1}F_0^2
\left(C_{udus}^{LLLL,S}+C_{udus}^{RRRR,S}\right),
\label{c5}
\\ %%
c_6&=&\frac{5}{3} g_{27\times1} F_0^2
\left(C_{udus}^{LLLL,P}+C_{udus}^{RRRR,P} \right).
\label{c6}
\end{eqnarray}
%%%%%
The spin-summed squared matrix element for the decay $K^-(k)\rightarrow\pi^+(p)l^-(q_1)l^-(q_2)$ is
\begin{eqnarray}
\nonumber
|\mathcal{M}|^2&=&
\frac{1}{2}\left|2 c_1+c_5 \left(m_K^2+m_\pi^2-s\right)\right|^2
\left(s-4 m_l^2\right)
+\frac{1}{2}\left|2 c_2+c_6\left(m_K^2+m_\pi^2-s\right)\right|^2 s
\\
\nonumber
&&+2\textrm{Re }\left\{\left[2 c_2+c_6\left(m_K^2+m_\pi^2-s\right)\right]
\left[\left(c_3^*-c_4^*\right)\left(m_K^2-m_\pi^2\right)
+\left(c_3^*+c_4^*\right)s\right]m_{l}\right\}
\\
&&+2c_4^*\left(\left(c_3+c_4\right)\left(s-m_K^2+m_\pi^2\right)
-2\left(c_3-c_4\right)\left(m_\pi^2+t\right)\right) m_l^2
\\\nonumber
&&+2c_3^*\left(\left(c_3+c_4\right)\left(s-m_\pi^2+m_K^2\right)
+2\left(c_3-c_4\right)\left(m_K^2+t\right)\right) m_l^2
\\\nonumber
&&-2\left|c_3-c_4 \right|^2\left(m_\pi^2 m_K^2+m_l^4-t\left(m_K^2+m_\pi^2-s-t\right)\right),
\end{eqnarray}
where $m_{K,\pi,l}$ are the masses of the $K^-$, $\pi^+$, and $l$ respectively, and $s=(q_1+q_2)^2$, $t=(p+q_2)^2$. The decay width is calculated as
\begin{eqnarray}
\Gamma=\frac{1}{2!}\frac{1}{2m_K}\frac{1}{128\pi^3m_K^2}\int ds \int dt~ |\mathcal{M}|^2,
\label{decaywidth}
\end{eqnarray}
where the integration domains are
\begin{eqnarray}
&&s\in \left[4m_l^2,~(m_K-m_\pi)^2\right],
\\
&&t\in \left[(E_2^*+E_3^*)^2-\left(\sqrt{E_2^{*2}-m_l^2}+\sqrt{E_3^{*2}-m_\pi^2}
\right)^2,~(E_2^*+E_3^*)^2-\left(\sqrt{E_2^{*2}-m_l^2}
-\sqrt{E_3^{*2}-m_\pi^2}\right)^2\right],
\end{eqnarray}
with
\begin{eqnarray}
E_2^*=\frac{1}{2}\sqrt{s},~E_3^*=\frac{1}{2}\frac{m_K^2-m_\pi^2-s}{\sqrt{s}}.
\end{eqnarray}

\begin{figure}
\centering
\includegraphics[width=11cm]{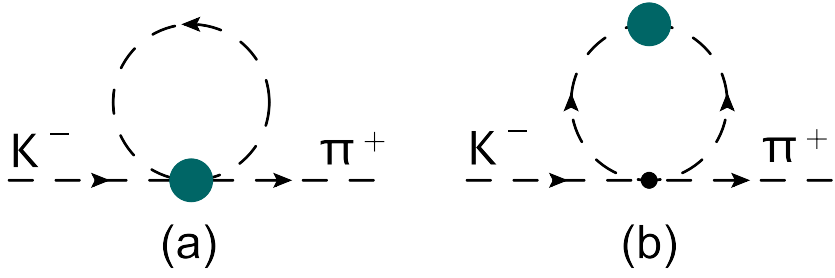}
\caption{Feynman diagrams for one-loop corrections to the $K^-\to\pi^+$ transition amplitude due to one insertion of hadronic operators (in blob) shown in table~\ref{tab3}. The small dot stands for a $\chi$PT vertex at $O(p^2)$.}
\label{fig3}
\end{figure}

We have been working so far at the first nonvanishing order of each operator in $\chi$PT. To estimate the errors incurred by ignoring higher-order terms we calculate the one-loop chiral logarithms contributing to the $K^-\rightarrow \pi^+$ transition amplitude
as shown by Feynman diagrams in figure~\ref{fig3}. These nonanalytic chiral logarithms cannot be cancelled by the counterterms to the one-loop diagrams, and thus can be used as a rough estimate of higher order terms. For this, we need the $K^-\pi^-|P|^2$ terms from expansion of the hadronic operators shown in table~\ref{tab3} with $P$ being any NGB, and the $K^+K^-\pi^+\pi^-$ terms in $\chi$PT at $O(p^2)$,
\begin{eqnarray}
\mathcal{L}^{\chi\textrm{PT}}_2&=&
\frac{F_0^2}{4}{\Tr}[\partial_\mu\Sigma\partial^\mu \Sigma^\dagger]
+\frac{F_0^2}{4}2B_0{\Tr}[M^\dagger\Sigma^\dagger+\Sigma M]
\nonumber
\\
&\supset&\frac{1}{6F_0^2}\Big[ (m_\pi^2+m_K^2)K^+\pi^+K^-\pi^-
+\partial^2(K^+\pi^+)K^-\pi^-
\nonumber
\\
&&+2 (\partial_\mu K^+)(\partial^\mu \pi^+)K^-\pi^-
+2 K^+\pi^+ (\partial_\mu K^-)(\partial^\mu \pi^-)\Big],
\end{eqnarray}
where $B_0$ is the order parameter for chiral symmetry breaking through the quark condensate, $-3F_0^2B_0=\langle0|\overline{q}q |0\rangle$. We will use the Gell-Mann-Okubo formula for NGBs, $3m_{\eta}^2=4m_K^2-m_\pi^2$.

For the purpose of illustrating the relative size of chiral logarithms to the leading terms, we consider the interactions from $g_{27\times 1}$, $g_{8\times8}^{a/b}$, and $g_{6\times\bar{6}}^{a/b}$ in table~\ref{tab3}. As will be clear in the next section, only the first two types of couplings can be obtained from matching to the SMEFT dim-7 operators. Since this discussion concerns only the accuracy of $\chi$PT at LO, we consider only the hadronic factors of those operators with a $(\checkmark)$ in table~\ref{tab3}, instead of a complete effective interaction, which is a sum involving various unknown Wilson coefficients and different lepton bilinears. Dropping the momentum squared of the leptonic system, $(k-p)^2=0$, we find
\begin{eqnarray}
\mathcal{M}_{27\times 1}&=&
\frac{5}{12}g_{27\times 1}F_K^2(m_K^2+m_\pi^2)\left[1-\frac{1}{4}
\left(\frac{17m_\pi^2-9m_K^2}{2(m_K^2-m_\pi^2)}L_\pi
-\frac{5m_K^2-m_\pi^2}{m_K^2-m_\pi^2}L_K+\frac{3}{2}L_\eta\right)\right],
\\
\mathcal{M}_{8\times 8}^{a/b}&=&
\frac{1}{2}g_{8\times 8}^{a/b}F_K^2\left[1
-\frac{1}{4}\left(\frac{9m_\pi^2-m_K^2}{2(m_K^2-m_\pi^2)}L_\pi
-\frac{m_K^2+3m_\pi^2}{m_K^2-m_\pi^2}L_K
+\frac{3}{2}L_\eta\right)\right],
\\
\mathcal{M}_{6\times \overline 6}^{a/b}&=&
\frac{1}{2}g_{6\times \overline 6}^{a/b} F_K^2\left[1
-\frac{1}{4}\left(\frac{9m_\pi^2-m_K^2}{2(m_K^2-m_\pi^2)}L_\pi
-\frac{5m_K^2-m_\pi^2}{m_K^2-m_\pi^2}L_K
+\frac{17}{6}L_\eta\right)\right],
\end{eqnarray}
where $L_P=m_P^2/(4\pi F_0)^2\ln(\mu^2/m_P^2)$ and $\mu$ is the renormalization scale. We have taken into account in the same approximation the renormalization of the decay constants~\cite{Gasser:1984gg}
\begin{eqnarray}
F_\pi&=&F_0\left[1+\frac{1}{2}\left(2L_\pi +L_K \right)\right],
\\
F_K&=&F_0\left[1+\frac{3}{8}\left(L_\pi +2L_K + L_\eta\right)\right],
\end{eqnarray}
and the wave function renormalization constants
\begin{eqnarray}
Z_\pi&=&1-\frac{1}{3}\left(2L_\pi + L_K \right),
\\
Z_K&=&1-\frac{1}{4}\left(L_\pi +2L_K + L_\eta\right).
\end{eqnarray}
The relative corrections in the three amplitudes have a magnitude of about $49.8\%,~41.6\%,~28.3\%$ ($34\%,~32.5\%,~31.3\%$) respectively, at the renormalization scale $\mu=\Lambda_\chi$ ($\mu=m_K$). This roughly fits the usual expectation on the accuracy of $SU(3)$ $\chi$PT.

\section{Matching to SMEFT}
\label{sec5}

Our previous results on short-distance contributions to the decay $K^-\to\pi^+l^-l^-$ are general in that they are not specific to physics above the electroweak scale $\Lambda_\textrm{EW}$ but rely only on well established symmetries at low energy. But to study the impact of low energy measurements on new physics at a high scale, we should connect our results to the effective field theory above $\Lambda_\textrm{EW}$. For this we make the minimal assumption that there are no new particles of mass of order $\Lambda_\textrm{EW}$ or smaller, so that SM appears as the leading terms in an EFT, i.e., SMEFT.

SMEFT is defined between $\Lambda_\textrm{EW}$ and some new physics scale $\Lambda_\textrm{NP}$ where new heavy particles have been integrated out. Its Lagrangian consists of a tower of effective operators built by the SM fields and satisfying the SM gauge symmetry $SU(3)_C\times SU(2)_L\times U(1)_Y$~\cite{Weinberg:1979sa,Buchmuller:1985jz,Grzadkowski:2010es,
Lehman:2014jma,Liao:2016hru,Liao:2019tep,Lehman:2015coa,Henning:2015alf,
Henning:2017fpj}:
\begin{eqnarray}
\mathcal{L}_{\text{SMEFT}}=\mathcal{L}_{\text{SM}}+\sum_{d>4,i}
C^d_i\calO^d_i,
\end{eqnarray}
where $\calO^d_i$ is the $i$-th operator of dimension $d$ with the corresponding Wilson coefficient $C^d_i$. These coefficients are treated independent in the EFT approach. But once a fundamental theory or another EFT at an even higher scale $\Lambda_\textrm{NP}$ is specified, the structure of the coefficients usually simplifies. This kind of simplification also manifests itself in the matching between SMEFT and LEFT as we will show below in the $|\Delta L|=2$ sector. In this work we will keep as general as possible concerning new physics but mentioning at the end of this section a few ultraviolet completions that yield SMEFT upon integrating out new heavy particles. In SMEFT lepton number violation first appears at dim-5 through the Weinberg operator that yields the Majorana neutrino mass upon electroweak symmetry breaking~\cite{Weinberg:1979sa}. This operator however is not directly relevant to our purpose of calculating short-distance contributions to the LNV $K^\pm$ decay. The leading contributions then arise from dim-7 operators in SMEFT. The effects of dim-7 operators in nuclear $0\nu\beta\beta$ decay have been analyzed recently by many groups, see, e.g., refs.~\cite{Cirigliano:2017djv,Cirigliano:2018yza,Liao:2019tep}
and references cited therein, and a few Wilson coefficients have been constrained from the experimental bounds~\cite{Albert:2014awa,KamLAND-Zen:2016pfg}. The Wilson coefficients to be constrained by the $K^\pm$ decay below are actually left free in nuclear $0\nu\beta\beta$ decay.

The SM particles gain mass when spontaneous electroweak symmetry breakdown takes place. By integrating out heavy particles like the weak gauge bosons $W^\pm,~Z$, the Higgs boson $h$, and the top quark, we match SMEFT to LEFT at the scale $\mu=\Lambda_\textrm{EW}$. The dim-9 operators in LEFT shown in equation~\eqref{ope_bas} arise from dim-7 operators combined with the SM weak gauge interactions in SMEFT. We find that only the operators $\calO_{udus}^{LLLL,S/P}$, $\tilde{\calO}_{udus}^{LRRL,S/P}$, and $\tilde{\calO}_{usud}^{LRRL,S/P}$ in LEFT are induced, with the coefficients,
\begin{eqnarray}
C_{udus}^{LLLL,S/P}&=&-2\sqrt{2}G_FV_{ud}V_{us}\left( C_{LHD1}^{ll\dagger}+4C_{LHW}^{ll\dagger}  \right),
\label{mat1}
\\
\tilde{C}_{udus}^{LRRL~S/P}&=&
-2\sqrt{2}G_FV_{us}C_{\bar{d}uLLD}^{11ll\dagger},
\label{mat2}
\\
\tilde{C}_{usud}^{LRRL~S/P}&=&
-2\sqrt{2}G_FV_{ud}C_{\bar{d}uLLD}^{21ll\dagger},
\label{mat3}
\end{eqnarray}
where both sides are evaluated at the scale $\mu=\Lambda_\textrm{EW}$, and $G_F$ is the Fermi constant, $V_{ud}$ and $V_{us}$ are the CKM matrix elements. On the right-hand side we have used the basis and notation of dim-7 operators in ref.~\cite{Liao:2016hru}; e.g., $C_{\bar{d}uLLD}^{21ll\dagger}$ refers to the $s,~u$ quarks and the lepton $l=e,~\mu$.

The above matching between LEFT and SMEFT simplifies considerably our subsequent phenomenological analysis. First of all, the QCD RGEs for the above coefficients decouple from other coefficients:
\begin{eqnarray}
\mu\frac{d}{d\mu}C_{udus}^{LLLL,S/P} &=&-\frac{\alpha_s}{2\pi}\left(\frac{3}{N}-3 \right)C_{udus}^{LLLL,S/P},
\\
\mu\frac{d}{d\mu}
\begin{pmatrix}
C_{uiuj}^{LRRL,S/P}
\\
\tilde{C}_{uiuj}^{LRRL,S/P}
\end{pmatrix}
&=&
-\frac{\alpha_s}{2\pi}
\begin{pmatrix}
6C_F & 3 \\
0 &  -\frac{3}{N}
\end{pmatrix}
\begin{pmatrix}
C_{uiuj}^{LRRL,S/P}
\\
\tilde{C}_{uiuj}^{LRRL,S/P}
\end{pmatrix},
\end{eqnarray}
where $(i,j)=(d,s),~(s,d)$. The solutions are
\begin{eqnarray}
C_{udus}^{LLLL,S/P}(\mu_1)&=&
\left[\frac{\alpha_s(\mu_2)}{\alpha_s(\mu_1)} \right]^{-\frac{2}{b}}
C_{udus}^{LLLL,S/P}(\mu_2),
\\
\tilde{C}_{uiuj}^{LRRL,S/P}(\mu_1)&=&
\left[\frac{\alpha_s(\mu_2)}{\alpha_s(\mu_1)} \right]^{-\frac{1}{b}}
\tilde{C}_{uiuj}^{LRRL,S/P}(\mu_2),
\\
C_{uiuj}^{LRRL,S/P}(\mu_1)&=&
\left[\frac{\alpha_s(\mu_2)}{\alpha_s(\mu_1)} \right]^{\frac{8}{b}}
C_{uiuj}^{LRRL,S/P}(\mu_2)
+\frac{1}{3}\left(\left[\frac{\alpha_s(\mu_2)}{\alpha_s(\mu_1)} \right]^{\frac{8}{b}}-\left[\frac{\alpha_s(\mu_2)}{\alpha_s(\mu_1)} \right]^{-\frac{1}{b}} \right)\tilde{C}_{uiuj}^{LRRL,S/P}(\mu_2),
\end{eqnarray}
where $b=-11+2n_f/3$ with $n_f$ being the number of active quark flavors between scales $\mu_1$ and $\mu_2$; in particular,
\begin{eqnarray}
C_{udus}^{LLLL,S/P}(\Lambda_\chi)&=&
0.78C_{udus}^{LLLL,S/P}(\Lambda_{\text{EW}}),
\label{rg1}
\\
\tilde{C}_{uiuj}^{LRRL,S/P}(\Lambda_\chi)&=&
0.88\tilde{C}_{uiuj}^{LRRL,S/P}(\Lambda_{\text{EW}}),
\label{rg2}
 \\
C_{uiuj}^{LRRL,S/P} (\Lambda_\chi)&=&
0.62\tilde{C}_{uiuj}^{LRRL,S/P}(\Lambda_{\text{EW}}).
\label{rg3}
\end{eqnarray}
The coefficients $c_i$ in equations~\eqref{c1}-\eqref{c6} for the $K$ decay simplify to
\begin{eqnarray}
\nonumber
c_1=c_2&=&g_{8\times 8}^{a}F_0^2
\left(C_{udus}^{LRRL,S/P}(\Lambda_\chi)
+C_{usud}^{LRRL,S/P}(\Lambda_\chi) \right)
\\
\nonumber
&&+g_{8\times 8}^{b}F_0^2
\left(\tilde{C}_{udus}^{LRRL,S/P}(\Lambda_\chi)
+\tilde{C}_{usud}^{LRRL,S/P}(\Lambda_\chi) \right)
\\
\nonumber
&=&\left(0.62 g_{8\times 8}^{a}+0.88g_{8\times 8}^{b}  \right)F_0^2
\left(\tilde{C}_{udus}^{LRRL,S/P}(\Lambda_{\text{EW}})
+\tilde{C}_{usud}^{LRRL,S/P}(\Lambda_{\text{EW}}) \right)
\\
&=&-2\sqrt{2}G_FV_{ud}V_{us}
\left(V_{ud}^{-1}C_{\bar{d}uLLD}^{11ll\dagger}(\Lambda_{\text{EW}})
+V_{us}^{-1}C_{\bar{d}uLLD}^{21ll\dagger}(\Lambda_{\text{EW}})\right)
\left(0.62 g_{8\times 8}^a+0.88g_{8\times 8}^b\right)F_0^2,
\label{smeftc1}
\\
c_3=c_4&=&0,
\label{smeftc2}
\\
\nonumber
c_5=c_6&=&\frac{5}{3}g_{27\times1}F_0^2C_{udus}^{LLLL,S/P}(\Lambda_\chi)
\\
\nonumber
&=&1.3g_{27\times1}
C_{udus}^{LLLL,S/P}(\Lambda_{\text{EW}})F_0^2
\\
&=&-2\sqrt{2}G_FV_{ud}V_{us}
\left( C_{LHD1}^{ll\dagger}(\Lambda_{\text{EW}})
+4C_{LHW}^{ll\dagger}(\Lambda_{\text{EW}})\right)
(1.3g_{27\times1})F_0^2,
\label{smeftc5}
\end{eqnarray}
and the squared matrix element becomes a compact form
\begin{eqnarray}
|\mathcal{M}_{\text{SMEFT}}|^2=\left|2 c_1+c_5 \left(m_K^2+m_\pi^2-s\right)\right|^2\left(s-2 m_l^2\right).
\label{smeftm2}
\end{eqnarray}

Normalizing the LNV $K$ decay width to its total width~\cite{Tanabashi:2018oca} yields the branching ratios for the decays to two identical leptons:
\begin{eqnarray}
\nonumber
\mathcal{B}^{\text{SMEFT}}_{K^-\rightarrow\pi^+\mu^-\mu^-}&=&
\Big\{8.1\times 10^{-4}\left|C_{\bar{d}uLLD}^{1122\dagger}
+4.4C_{\bar{d}uLLD}^{2122\dagger}\right|^2
+6.7\times 10^{-8}\left|C_{LHD1}^{22\dagger}+4C_{LHW}^{22\dagger}\right|^2
\\
&&+1.5\times 10^{-5} \textrm{Re~}\left[ \left(C_{\bar{d}uLLD}^{1122\dagger}
+4.4C_{\bar{d}uLLD}^{2122\dagger}\right) \left(C_{LHD1}^{22}+4C_{LHW}^{22}\right)\right]\Big\}\GeV^6,
\\
\nonumber
\mathcal{B}^{\text{SMEFT}}_{K^-\rightarrow\pi^+e^-e^-}&=&
\Big\{2.3\times 10^{-3}\left|C_{\bar{d}uLLD}^{1111\dagger}
+4.4C_{\bar{d}uLLD}^{2111\dagger}\right|^2
+2.3\times 10^{-7}\left|C_{LHD1}^{11\dagger}+4C_{LHW}^{11\dagger}\right|^2
\\
&&+4.6\times 10^{-5}\textrm{Re~}\left[ \left(C_{\bar{d}uLLD}^{1111\dagger}
+4.4C_{\bar{d}uLLD}^{2111\dagger}\right) \left(C_{LHD1}^{11}+4C_{LHW}^{11}\right) \right]\Big\}\GeV^6.
\end{eqnarray}
To get some feel about the bound on the relevant energy scale, we assume naively that the above Wilson coefficients in SMEFT scale as $\Lambda^{-3}$, then the experimental upper bounds in table~\ref{tab2} translate into a loose bound $\Lambda> \calO(10~\GeV)$. This bound is indeed much weaker than that from nuclear $0\nu\beta\beta$ decay, $\Lambda> \calO(10~\TeV)$~\cite{Liao:2019tep}, but it concerns the quarks and leptons of the second generation. This relative weakness arises largely from much smaller data samples in $K$ decays than in nuclear $0\nu\beta\beta$ decay: while about $10^{11}$ $K^+$ particles were used to search for each of the above two decays in the NA62 experiment~\cite{CortinaGil:2019dnd}, there are about $10^{27}$ $^{136}\textrm{Xe}$ in the KamLAND-Zen experiment~\cite{KamLAND-Zen:2016pfg}. Taking the sixth root of the ratio of the numbers of particles involved already accounts for about a half-thousand difference in the lower bounds that can be respectively set on $\Lambda$. One should be careful in interpreting the above loose bound. It does not mean that the EFT approach here would be valid for a new physics scale above $\calO(10~\GeV)$, which indeed cannot be the case, but suggests that an NA62-type of experiment cannot yield a useful bound if the relevant effective scale involved in the second generation of fermions is as high as the one reached in experiments of nuclear $0\nu\beta\beta$ decay involving only the first generation of fermions.

\begin{figure}
\centering
\includegraphics[width=12cm]{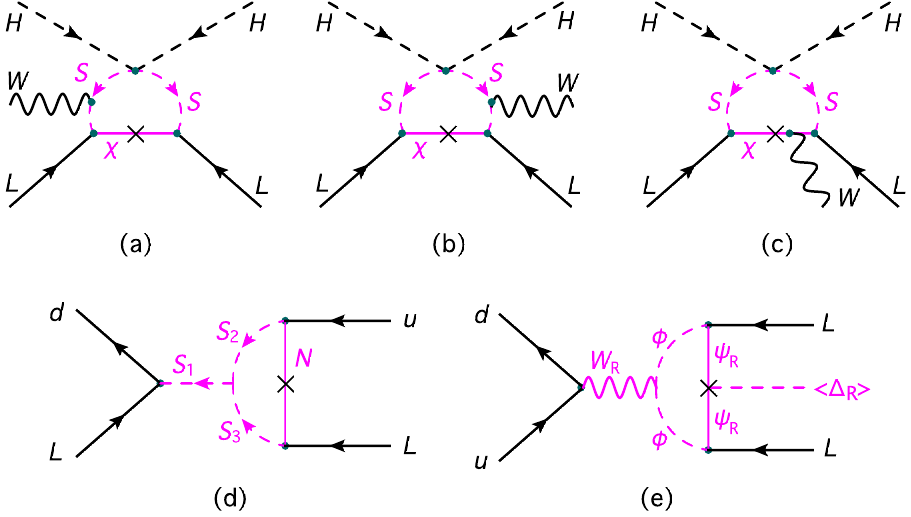}
\caption{Feynman diagrams for generating operator $\calO_{LHW}$ (a-c) in color octet models and operator $\calO_{\bar{d}uLLD}$ in leptoquark models (d) and left-right symmetric models (e). New heavy particles are highlighted in magenta, and $\times$ stands for a chirality flip.}
\label{fig4}
\end{figure}

Before ending this section we discuss briefly some typical new physics realizations of the relevant operators $\calO_{LHW},~\calO_{LHD1},~\calO_{\bar{d}uLLD}$ in SMEFT upon integrating out heavy particles. The operator $\calO_{LHW}$, relevant to neutrino electromagnetic transition moments, is generated in models of radiative neutrino mass~\cite{Ma:1998dn,Bonnet:2012kz,Sierra:2014rxa,Cepedello:2017eqf,Bonnet:2009ej,Cai:2017jrq} by attaching an $SU(2)_L$ gauge boson to corresponding neutrino mass diagrams. An example is shown in figure~\ref{fig4}(a-c) for the color octet neutrino mass model~\cite{FileviezPerez:2009ud}, where $S$ and $\chi$ are the color octet scalar and fermion with the SM quantum numbers $(8,2,1/2)$ and $(8,3,0)$, respectively. The operator $\calO_{LHD1}$ can be similarly generated by keeping quadratic momentum terms from neutrino mass diagrams. The operator $\calO_{\bar{d}uLLD}$ involves both quarks and leptons, and can thus be most easily realized in leptoquark models; see, for instance, Ref.~\cite{Dorsner:2016wpm} for a review. We show an example in figure~\ref{fig4}(d), where the leptoquarks $S_1,~S_2$ have the quantum numbers $(3,2,1/6),~(3,1,2/3)$ and the scalar doublet $S_3$ and fermion singlet $N$ have $(1,2,-1/2),~(1,1,0)$, respectively. The potential mixing of $S_3$ with the SM Higgs may be avoided if necessary by assigning an odd $Z_2$ parity to all of $S_2,~S_3,~N$. The gauge covariant derivative for the operator is formed by attaching a gauge boson line in figure~\ref{fig4}(d). This operator can also be realized in left-right symmetric models~\cite{Mohapatra:1979ia,Mohapatra:1980yp} which have an extended gauge group $SU(2)_L\times SU(2)_R\times U(1)_{B-L}$. A concrete example is shown in figure~\ref{fig4}(e). Here $\langle\Delta_R\rangle$ stands for the large vacuum expectation value of the scalar triplet $\Delta_R$ which triggers the symmetry breakdown $SU(2)_R\times U(1)_{B-L}\to U(1)_Y$, and $\psi_R=(N,e)$ and $W_R$ the right-handed lepton doublet and gauge bosons of $SU(2)_R$ respectively. Upon symmetry breaking both $N$ and $W_R$ gain a large mass, and the bidoublet scalar $\phi$ splits into an SM-like Higgs part and a heavy part~\cite{Maiezza:2016ybz}. Note that only the heavy scalar part of $\phi$ appears in figure~\ref{fig4}(e), and for clarity we have shown the components $u,~d$ of the right-handed quark doublet. Our above analysis fits well the classification of the tree- and loop-generated operators in Ref.~\cite{Einhorn:2013kja}.

%%%%%
\section{Conclusion}
\label{sec6}

The Majorana nature of neutrinos has so far been intensively explored in nuclear $0\nu\beta\beta$ decay both experimentally and theoretically. Considering the null result in current experimental searches it is important that we seek other potential signals that could reveal the Majorana nature of neutrinos. The LNV decays of the charged mesons and $\tau$ lepton may play a role in probing interactions of heavier quarks and leptons to which nuclear $0\nu\beta\beta$ decay is not sensitive.

In this work we have for the first time investigated the LNV decay $K^-\rightarrow\pi^+l^-l^-$ in the framework of effective field theory. We established the basis of $|\Delta L|=2$ dim-9 operators in LEFT that are responsible for the leading order short-distance contributions to LNV processes including, e.g., the decays of the mesons and $\tau$ lepton and nuclear $0\nu\beta\beta$ decay. We calculated the one-loop QCD RGEs for the basis operators and provided both analytical and numerical solutions. Then by restricting ourselves to the three light quarks, we matched the above effective interactions in LEFT to those in $\chi$PT for the octet pseudoscalar mesons at the first nonvanishing order of each operator. We made a complete analysis on the irreducible representations of the operators under chiral symmetries and introduced hadronic low energy constants for independent structures. We also estimated errors due to neglect of higher orders in $\chi$PT by computing chiral logarithms for a few transition amplitudes, and found their size is consistent with the usual expectation. Our results on the short-distance contributions to the decay $K^-\rightarrow\pi^+l^-l^-$ are general in that they are based only on QED, QCD, and chiral symmetries at low energy, are parameterized by Wilson coefficients in LEFT and LECs in $\chi$PT, but do not depend on dynamical details at high energy scales. To connect to new physics at a high scale, we matched our results in LEFT to SMEFT at the electroweak scale $\Lambda_\textrm{EW}$ assuming that there are no new particles with a mass of order $\Lambda_\textrm{EW}$ or below. We found that this simplifies the structures of LEFT that enter in the $K^-$ decay width. In this manner we can translate the experimental bounds on low energy processes to those on the Wilson coefficients in SMEFT, stringing all the way the series of EFTs from SMEFT to LEFT and $\chi$PT. Parameterizing the coefficients roughly by $\sim\Lambda^{-3}$, the current bounds on LNV $K$ decays yield a loose bound $\Lambda> \calO(10~\GeV)$. Nevertheless, this is the first bound based exclusively on EFTs in the second generation of quarks and leptons.

%\newpage
\vspace{0.5cm}
\noindent %
\section*{Acknowledgement}

This work was supported in part by the Grants No.~NSFC-11975130, No.~NSFC-11575089, No.~NSFC-11025525, by The National Key Research and Development Program of China under Grant No. 2017YFA0402200, by the CAS Center for Excellence in Particle Physics (CCEPP), and by the Grant No.~MOST~106-2112-M-002-003-MY3. We thank Feng-Kun Guo for helpful discussions.

%\newpage

\begin{appendices}
\numberwithin{equation}{section}
\setcounter{equation}{0}

%%%%%
\section{Redundant LNV dim-9 operators in LEFT}
\label{app1}

In this appendix, we will detail our determination of the basis for the dim-9 operators with $|\Delta L|=2$, and express redundant operators as linear combinations of the basis operators in table~\ref{tab1} by using various Fierz identities. For brevity, we use $\Gamma_1\otimes \Gamma_2$ and $\Gamma_1\odot \Gamma_2$ to denote the four-fermion operator $(\overline{\Psi_1} \Gamma_1 \Psi_2)\otimes (\overline{\Psi_3} \Gamma_2 \Psi_4)$ and its corresponding Fierz transformed partner $(\overline{\Psi_1} \Gamma_1 \Psi_4)\odot (\overline{\Psi_3} \Gamma_2 \Psi_2)$. In section~\ref{sec2}, all operators are written as a product of a four-quark factor and a lepton factor. Based on this, now we show that all other operators involving various $\Gamma_i$ structures are redundant.

First of all, from the properties of Dirac matrices, we have the following Fierz identities
\begin{eqnarray}
\frac{i}{2}\epsilon^{\mu\nu\rho\sigma}\sigma_{\rho\sigma}&=&
\sigma^{\mu\nu}\gamma_5~~(\epsilon^{0123}=+1),
\\
\sigma^{\mu\nu} P_{\pm}\otimes \sigma_{\mu\nu} P_{\mp}&=&0,
\\
\sigma^{\mu\rho} P_{\pm}\otimes\sigma^{\nu}_{~~\rho}P_{\mp}\times j^{\alpha\beta}_{\mu\nu}&=&0,
\\
\gamma^\mu P_{\pm}\otimes\gamma^\nu P_{\pm}\times j_{5\mu\nu}^{\alpha\beta}&=&\mp\gamma^\mu P_{\pm}\odot \gamma^\nu P_{\pm}\times j^{\alpha\beta}_{\mu\nu},
\\
\sigma^{\mu\nu} P_{\pm}\otimes P_{\pm/\mp}\times j_{5\mu\nu}^{\alpha\beta}&=&
\pm\sigma^{\mu\nu} P_{\pm}\otimes P_{\pm/\mp}\times j^{\alpha\beta}_{\mu\nu},
\\
\sigma^{\mu\rho} P_{\pm}\otimes \sigma^{\nu}_{~\rho} P_{\pm}\times j_{5\mu\nu}^{\alpha\beta}&=&
\pm\sigma^{\mu\rho} P_{\pm}\otimes \sigma^{\nu}_{~~\rho} P_{\mp}\times j^{\alpha\beta}_{\mu\nu},
\end{eqnarray}
where $j_{5\mu\nu}^{\alpha\beta}=
\overline{l^\alpha}\gamma_5\sigma_{\mu\nu}l^{\beta,C}$ and $P_\pm=(1\pm\gamma_5)/2$. Thus we can discard the lepton bilinear $j_{5\mu\nu}^{\alpha\beta}$ together with the tensor $\epsilon^{\mu\nu\rho\sigma}$ for contraction between the quark and lepton factors and include only lepton bilinears listed in table~\ref{tab1}.
\begin{itemize}
%%%
\item Type-$(\overline{u_L^p}\Gamma_1 d_L^r)(\overline{u_L^s}\Gamma_1 d_L^t)(\overline{l_\alpha}\Gamma_3 l_\beta^C)$
    \\
Redundant operator:
\begin{eqnarray}
(\overline{u_L^p}\gamma^\mu d_L^r][\overline{u_L^s}\gamma_\mu d_L^t)( j^{\alpha\beta}/j_5^{\alpha\beta})=(\overline{u_L^p}\gamma^\mu d_L^t)[\overline{u_L^s}\gamma_\mu d_L^r]( j^{\alpha\beta}/j_5^{\alpha\beta})=\calO_{ptsr}^{LLLL,S/P},
\end{eqnarray}
by the Fierz identity
\begin{eqnarray}
\gamma^\mu P_{\pm}\otimes \gamma_\mu P_{\pm}=\gamma^\mu P_{\pm}\odot \gamma_\mu P_{\pm}.
\end{eqnarray}
%%%
\item Type-$(\overline{u_L}\Gamma_1 d_R)(\overline{u_L}\Gamma_1 d_R)(\overline{l_\alpha}\Gamma_3 l_\beta^C)$\\
Redundant operators:
\begin{eqnarray}
(\overline{u_L^p}\sigma^{\mu\nu}d_R^r)
[\overline{u_L^s}\sigma_{\mu\nu}d_R^t](j^{\alpha\beta}/j_5^{\alpha\beta})
&=&-4\calO_{prst}^{LRLR,S/P}-8\tilde{\calO}_{ptsr}^{LRLR,S/P},
\\
(\overline{u_L^p}\sigma^{\mu\nu}d_R^r]
[\overline{u_L^s}\sigma_{\mu\nu}d_R^t)( j^{\alpha\beta}/j_5^{\alpha\beta})
&=&-4\tilde{\calO}_{prst}^{LRLR,S/P}-8\calO_{ptsr}^{LRLR,S/P},
\\
\nonumber
(\overline{u_L^p}i\sigma^{\mu\nu}d_R^r]
[\overline{u_L^s}d_R^t)(j_{\mu\nu}^{\alpha\beta})&=&
-\frac{1}{2}(\overline{u_L^p}i\sigma^{\mu\nu}d_R^t)
[\overline{u_L^s}d_R^r](j_{\mu\nu}^{\alpha\beta})
-\frac{1}{2}(\overline{u_L^p}d_R^t)
[\overline{u_L^s}i\sigma^{\mu\nu}d_R^r](j_{\mu\nu}^{\alpha\beta})
\\
\nonumber
&&
+\frac{1}{2}(\overline{u_L^p}\sigma^{\mu}_{~~\rho}d_R^t)
[\overline{u_L^s}\sigma^{\nu\rho} d_R^r](j_{\mu\nu}^{\alpha\beta})
\\
&=&-\frac{1}{2}\calO_{ptsr}^{LRLR,T}
-\frac{1}{2}\calO_{srpt}^{LRLR,T}
+\frac{1}{2}\tilde{\calO}_{ptsr}^{LRLR,T},
\\
\nonumber
(\overline{u_L^p}\sigma^{\mu\rho}d_R^r]
[\overline{u_L^s}\sigma^{\nu}_{~~\rho}d_R^t)(j_{\mu\nu}^{\alpha\beta})
&=&(\overline{u_L^p}i\sigma^{\mu\nu}d_R^t)
[\overline{u_L^s}d_R^r](j_{\mu\nu}^{\alpha\beta})
-(\overline{u_L^p}d_R^t)[\overline{u_L^s}i\sigma^{\mu\nu}d_R^r]
(j_{\mu\nu}^{\alpha\beta})
\\
&=&\calO_{ptsr}^{LRLR,T}-\calO_{srpt}^{LRLR,T},
\end{eqnarray}
by the Fierz identities
\begin{eqnarray}
\sigma^{\mu\nu} P_{\pm}\otimes \sigma_{\mu\nu} P_{\pm}&=&-4P_{\pm}\otimes  P_{\pm}-8 P_{\pm}\odot  P_{\pm},
\\
\nonumber
\sigma^{\mu\nu}P_{\pm}\otimes P_{\pm}&=&-\frac{1}{2}\Big( \sigma^{\mu\nu} P_{\pm}\odot P_{\pm}+P_{\pm}\odot \sigma^{\mu\nu}P_{\pm}\Big)
\\
&&-\frac{i}{4}\Big(\sigma^{\mu\rho}P_{\pm}\odot\sigma^{\nu}_{~~\rho}P_{\pm}
-\sigma^{\nu\rho}P_{\pm}\odot\sigma^{\mu}_{~~\rho}P_{\pm}\Big),
\\
\sigma^{\mu\rho} P_{\pm}\otimes\sigma^{\nu}_{~~\rho} P_{\pm}\times j^{\alpha\beta}_{\mu\nu}&=&
\left( i\sigma^{\mu\nu}P_{\pm}\odot P_{\pm}
-i P_{\pm}\odot\sigma^{\mu\nu} P_{\pm}\right)\times j^{\alpha\beta}_{\mu\nu}.
\end{eqnarray}
%%%
\item Type-$(\overline{u_L}\Gamma_1 d_R)(\overline{u_L}\Gamma_1 d_L)(\overline{l_\alpha}\Gamma_3 l_\beta^C)$
    \\
Redundant operators:
\begin{eqnarray}
(\overline{u_L^p}i\sigma^{\mu\nu}d_R^r)[\overline{u_L^s}\gamma_\mu d_L^t](j_\nu^{\alpha\beta} /j_{5\nu}^{\alpha\beta})&=&
-\calO_{prst}^{LRLL,V/A}-2\tilde{\calO}_{srpt}^{LRLL,V/A},
\\
(\overline{u_L^p}i\sigma^{\mu\nu}d_R^r][\overline{u_L^s}\gamma_\mu d_L^t)(j_\nu^{\alpha\beta} /j_{5\nu}^{\alpha\beta})&=&
-\tilde{\calO}_{prst}^{LRLL,V/A}-2\calO_{srpt}^{LRLL,V/A},
\end{eqnarray}
by the following Fierz identity
\begin{eqnarray}
 \sigma^{\mu\nu} P_\mp\otimes\gamma_\nu P_\pm=-iP_\mp \otimes \gamma^\mu P_\pm  -2i\gamma^\mu P_\pm \odot  P_\mp.
\end{eqnarray}
%%%
\item Type-$(\overline{u_L}\Gamma_1 d_R)(\overline{u_R}\Gamma_1 d_R)(\overline{l_\alpha}\Gamma_3 l_\beta^C)$
    \\
Redundant operators:
\begin{eqnarray}
(\overline{u_L^p}i\sigma^{\mu\nu}d_R^r)[\overline{u_R^s}\gamma_\mu d_R^t](j_\nu^{\alpha\beta} /j_{5\nu}^{\alpha\beta})&=&
\calO_{prst}^{LRRR,V/A}+2\tilde{\calO}_{ptsr}^{LRRR,V/A},
\\
(\overline{u_L^p}i\sigma^{\mu\nu}d_R^r][\overline{u_R^s}\gamma_\mu d_R^t)(j_\nu^{\alpha\beta} /j_{5\nu}^{\alpha\beta})&=&
\tilde{\calO}_{prst}^{LRRR,V/A}+2\calO_{ptsr}^{LRRR,V/A},
\end{eqnarray}
by the following Fierz identity
\begin{eqnarray}
\sigma^{\mu\nu}P_\pm \otimes  \gamma_\nu P_\pm =iP_\pm \otimes  \gamma^\mu P_\pm+2i P_\pm \odot  \gamma^\mu P_\pm.
\end{eqnarray}
%%%
\item Type-$(\overline{u_L}\Gamma_1 d_R)(\overline{u_R}\Gamma_1 d_L)(\overline{l_\alpha}\Gamma_3 l_\beta^C)$
    \\
Redundant operators
\begin{eqnarray}
(\overline{u_L^p}\gamma^\mu d_L^r)
[\overline{u_R^s}\gamma_\mu d_R^t](j^{\alpha\beta}/j_5^{\alpha\beta})
&=&-2\tilde{\calO}_{ptsr}^{LRRL,S/P},
\\
(\overline{u_L^p}\gamma^\mu d_L^r]
[\overline{u_R^s}\gamma_\mu d_R^t)(j^{\alpha\beta}/j_5^{\alpha\beta})
&=&-2\calO_{ptsr}^{LRRL,S/P},
\\
(\overline{u_L^p}\gamma^\mu d_L^r)
[\overline{u_R^s}\gamma^\nu d_R^t]( j_{\mu\nu}^{\alpha\beta})
&=&\frac{1}{2}\tilde{\calO}_{ptsr}^{LRRL,T}
-\frac{1}{2}\tilde{\calO}_{srpt}^{RLLR,T},
\\
(\overline{u_L^p}\gamma^\mu d_L^r]
[\overline{u_R^s}\gamma^\nu d_R^t)(j_{\mu\nu}^{\alpha\beta})
&=&\frac{1}{2}\calO_{ptsr}^{LRRL,T}
-\frac{1}{2}\calO_{srpt}^{RLLR,T},
\end{eqnarray}
by the following Fierz identities
\begin{eqnarray}
\gamma^\mu P_{\pm}\otimes \gamma_\mu P_{\mp}
&=& -2 P_{\mp}\odot P_{\pm},
\\
\gamma^\mu P_{\pm}\otimes \gamma^\nu P_{\mp}\times j^{\alpha\beta}_{\mu\nu}
&=&\frac{i}{2}\Big( \sigma^{\mu\nu}P_{\mp}\odot P_{\pm}-P_{\mp}\odot \sigma^{\mu\nu}P_{\pm}\Big)\times j^{\alpha\beta}_{\mu\nu}.
\end{eqnarray}
\end{itemize}

%%%%%
\section{Solutions to RGEs for dim-9 LNV operators in LEFT}
\label{app12}

In this Appendix we solve the complete set of one-loop QCD RGEs~\eqref{lrgi}-\eqref{lrgf}. Our results in previous sections on nuclear $0\nu\beta\beta$ decay and the decays $K^\pm\to\pi^\mp l_\alpha^\pm l_\beta^\pm$ form subsets of the results recorded below.

We denote the diagonal matrices formed with the eigenvalues of anomalous dimension matrices in equations~\eqref{lrgi}-\eqref{lrgf}:
\begin{eqnarray}
R_1&=&{\diag}\left(\zeta_{2/1}^{\frac{4}{3b}},~
\zeta_{2/1}^{-\frac{2}{3b}},\right),
\\
R_2&=&{\diag}\left(\zeta_{2/1}^{\frac{17+\sqrt{241}}{6b}},~
\zeta_{2/1}^{-\frac{\sqrt{241}+1}{6b}},~
\zeta_{2/1}^{\frac{{\sqrt{241}-1}}{6b}},~
\zeta_{2/1}^{\frac{17-\sqrt{241}}{6b}}\right),
\\
R_3&=&{\diag}\left(\zeta_{2/1}^{\frac{4}{b}},~
\zeta_{2/1}^{\frac{10}{3b}},~\zeta_{2/1}^{\frac{10}{3b}},~
\zeta_{2/1}^{-\frac{8}{3b}},~1\right),
\\
R_4&=&{\diag}\left(\zeta_{2/1}^{\frac{8}{3b}},~
\zeta_{2/1}^{-\frac{1}{3b}}\right),
\end{eqnarray}
where $\zeta_{2/1}=\alpha_s(\mu_2)/\alpha_s(\mu_1)$ and $b=-11+2n_f/3$ with $n_f$ being the number of active quarks between the scales $\mu_1$ and $\mu_2$. The corresponding diagonalization matrices are found to be,
\begin{eqnarray}
T_1&=&
\begin{pmatrix}
-1 & 1 \\
1 &  1
\end{pmatrix},
\\
T_2&=&
\begin{pmatrix}
\frac{21+\sqrt{241}}{10} & \frac{\sqrt{241}-15}{2}  & -\frac{\sqrt{241}+15}{2} & \frac{21-\sqrt{241}}{10} \\
-\frac{21+\sqrt{241}}{10} & \frac{\sqrt{241}-15}{2}  & -\frac{\sqrt{241}+15}{2} & -\frac{21-\sqrt{241}}{10} \\
-1 & 1 &  1 & -1\\
1 & 1 &  1 & 1
\end{pmatrix},
\\
T_3&=&
\begin{pmatrix}
1 & 0 & -1 & 0 & 1 \\
-1 & -6 & 0 & 0 & 1 \\
-1 & 6 & 0 & 0 & 1 \\
1 & 0 & 1 & 0 & 1 \\
0 & 1 & 0 & 1 & 0 \\
\end{pmatrix},
\\
T_4&=&
\begin{pmatrix}
1 & -1 \\
0 &  3
\end{pmatrix}.
\end{eqnarray}
Then the solutions to equations~\eqref{lrgi}-\eqref{lrgf} between the scales $\mu_1$ and $\mu_2$ are
\begin{eqnarray}
\begin{pmatrix}
C_{prst}^{LLLL,S/P}
\\
C_{ptsr}^{LLLL,S/P}
\end{pmatrix}
(\mu_1)
&=&
T_1R_1^3T_1^{-1}
\begin{pmatrix}
C_{prst}^{LLLL,S/P}
\\
C_{ptsr}^{LLLL,S/P}
\end{pmatrix}
(\mu_2),~
\\
\begin{pmatrix}
C_{prst}^{LLLL,T}
\\
\tilde{C}_{prst}^{LLLL,T}
\end{pmatrix}
(\mu_1)
&=&
T_1R_1T_1^{-1}
\begin{pmatrix}
C_{prst}^{LLLL,T}
\\
\tilde{C}_{prst}^{LLLL,T}
\end{pmatrix}
(\mu_2),
\\%%%
\begin{pmatrix}
C_{prst}^{LRLR,S/P}
\\
C_{ptsr}^{LRLR,S/P}
\\
\tilde{C}_{prst}^{LRLR,S/P}
\\
\tilde{C}_{ptsr}^{LRLR,S/P}
\end{pmatrix}
(\mu_1)
&=&
T_2R_2^2T_2^{-1}
\begin{pmatrix}
C_{prst}^{LRLR,S/P}
\\
C_{ptsr}^{LRLR,S/P}
\\
\tilde{C}_{prst}^{LRLR,S/P}
\\
\tilde{C}_{ptsr}^{LRLR,S/P}
\end{pmatrix}
(\mu_2),~
\\
%%%
\begin{pmatrix}
C_{prst}^{LRLR,T}
\\
C_{ptsr}^{LRLR,T}
\\
C_{srpt}^{LRLR,T}
\\
C_{stpr}^{LRLR,T}
\\
\tilde{C}_{prst}^{LRLR,T}
\end{pmatrix}
(\mu_1)
&=&
T_3R_3T_3^{-1}
\begin{pmatrix}
C_{prst}^{LRLR,T}
\\
C_{ptsr}^{LRLR,T}
\\
C_{srpt}^{LRLR,T}
\\
C_{stpr}^{LRLR,T}
\\
\tilde{C}_{prst}^{LRLR,T}
\end{pmatrix}
(\mu_2),
\\%%%
\begin{pmatrix}
C_{prst}^{LRLL,V/A}
\\
C_{srpt}^{LRLL,V/A}
\\
\tilde{C}_{prst}^{LRLL,V/A}
\\
\tilde{C}_{srpt}^{LRLL,V/A}
\end{pmatrix}
(\mu_1)
&=&
T_2R_2T_2^{-1}
\begin{pmatrix}
C_{prst}^{LRLL,V/A}
\\
C_{srpt}^{LRLL,V/A}
\\
\tilde{C}_{prst}^{LRLL,V/A}
\\
\tilde{C}_{srpt}^{LRLL,V/A}
\end{pmatrix}
(\mu_2),~
\\
%%%
\begin{pmatrix}
C_{prst}^{LRRR,V/A}
\\
C_{ptsr}^{LRRR,V/A}
\\
\tilde{C}_{prst}^{LRRR,V/A}
\\
\tilde{C}_{ptsr}^{LRRR,V/A}
\end{pmatrix}
(\mu_1)
&=&
T_2R_2T_2^{-1}
\begin{pmatrix}
C_{prst}^{LRRR,V/A}
\\
C_{ptsr}^{LRRR,V/A}
\\
\tilde{C}_{prst}^{LRRR,V/A}
\\
\tilde{C}_{ptsr}^{LRRR,V/A}
\end{pmatrix}
(\mu_2),
\\ %%%
\begin{pmatrix}
C_{prst}^{LRRL,T}
\\
\tilde{C}_{ptsr}^{LRRL,T}
\end{pmatrix}
(\mu_1)
&=&
T_4R_4T_4^{-1}
\begin{pmatrix}
C_{prst}^{LRRL,T}
\\
\tilde{C}_{ptsr}^{LRRL,T}
\end{pmatrix}
(\mu_2),~
\\
%%%
\begin{pmatrix}
C_{prst}^{LRRL,S/P}
\\
\tilde{C}_{ptsr}^{LRRL,S/P}
\end{pmatrix}
(\mu_1)
&=&
T_4R_4^3T_4^{-1}
\begin{pmatrix}
C_{prst}^{LRRL,S/P}
\\
\tilde{C}_{ptsr}^{LRRL,S/P}
\end{pmatrix}
(\mu_2).
\end{eqnarray}
For practical applications, we show the numerical results between the scale $\Lambda_\chi$ (on the left hand side) and the scale $\Lambda_{\text{EW}}$ (on the right hand side, not displayed for brevity) where quark threshold effects have been incorporated:
\begin{eqnarray}
C_{prst}^{LLLL,S/P}(\Lambda_\chi)&=&
1.22 C_{prst}^{LLLL,S/P}-0.44 C_{ptsr}^{LLLL,S/P},
\\
C_{prst}^{LLLL,T}(\Lambda_\chi)&=&
1.05 C_{prst}^{LLLL,T}-0.13\tilde{C}_{prst}^{LRLR,T},
\\
\tilde{C}_{prst}^{LLLL,T}(\Lambda_\chi)&=&
1.05 \tilde{C}_{prst}^{LLLL,T}-0.13C_{prst}^{LRLR,T},
\\
C_{prst}^{LRLR,S/P}(\Lambda_\chi)&=&
3.12 C_{prst}^{LRLR,S/P}-1.3C_{ptsr}^{LRLR,S/P}
+0.75\tilde{C}_{prst}^{LRLR,S/P}
-1.09 \tilde{C}_{ptsr}^{LRLR,S/P},
\\
\tilde{C}_{prst}^{LRLR,S/P}(\Lambda_\chi)&=&
0.54 \tilde{C}_{prst}^{LRLR,S/P} -0.02\tilde{C}_{ptsr}^{LRLR,S/P}
-0.5 C_{prst}^{LRLR,S/P}+0.42 C_{ptsr}^{LRLR,S/P},
\\
C_{prst}^{LRLR,T}(\Lambda_\chi)&=&
1.43C_{prst}^{LRLR,T}
-0.16\big(C_{ptsr}^{LRLR,T}+C_{srpt}^{LRLR,T}\big)
-0.1 C_{stpr}^{LRLR,T},
\\
\tilde{C}_{prst}^{LRLR,T}(\Lambda_\chi)&=&
0.71 \tilde{C}_{prst}^{LRLR,T}
-0.07\big(C_{ptsr}^{LRLR,T}-C_{srpt}^{LRLR,T}\big),
\\
C_{prst}^{LRLL,V/A}(\Lambda_\chi)&=&
1.75 C_{prst}^{LRLL,V/A} -0.4C_{srpt}^{LRLL,V/A}
+0.22\tilde{C}_{prst}^{LRLL,V/A}
-0.39\tilde{C}_{srpt}^{LRLL,V/A},
\\
\tilde{C}_{prst}^{LRLL,V/A}(\Lambda_\chi)&=&
0.79 \tilde{C}_{prst}^{LRLL,V/A}-0.07 \tilde{C}_{srpt}^{LRLL,V/A}
-0.17 C_{prst}^{LRLL,V/A}+0.13C_{srpt}^{LRLL,V/A},
\\
C_{prst}^{LRRR,V/A}(\Lambda_\chi)&=&
1.75 C_{prst}^{LRRR,V/A} -0.4C_{ptsr}^{LRRR,V/A}
+0.22\tilde{C}_{prst}^{LRRR,V/A}
-0.39\tilde{C}_{ptsr}^{LRRR,V/A},
\\
\tilde{C}_{prst}^{LRRR,V/A}(\Lambda_\chi)&=&
0.79 \tilde{C}_{prst}^{LRRR,V/A}-0.07 \tilde{C}_{ptsr}^{LRRR,V/A}
-0.17 C_{prst}^{LRRR,V/A}+0.13C_{ptsr}^{LRRR,V/A},
\\
C_{prst}^{LRRL,T}(\Lambda_\chi)&=&
1.4C_{prst}^{LRRL,T}+0.15\tilde{C}_{ptsr}^{LRRL,T},
\\
\tilde{C}_{prst}^{LRRL,T}(\Lambda_\chi)&=&
0.96\tilde{C}_{prst}^{LRRL,T},
\\
C_{prst}^{LRRL,S/P}(\Lambda_\chi)&=&
2.74C_{prst}^{LRRL,S/P}+0.62\tilde{C}_{ptsr}^{LRRL,S/P},
\\
\tilde{C}_{prst}^{LRRL,S/P}(\Lambda_\chi)&=&
0.88\tilde{C}_{prst}^{LRRL,S/P}.
\end{eqnarray}

%%%%%
\section{Basis of dim-7 operators in SMEFT}
\label{app2}

This appendix reproduces for completeness the basis of dim-7 operators in SMEFT that was obtained in ref.~\cite{Liao:2016hru}. The convention for fields in table~\ref{tab4} are as follows: $L,~Q$ are the left-handed lepton and quark doublet fields, $u,~d,~e$ are the right-handed up-type quark, down-type quark and charged lepton singlet fields, and $H$ denotes the Higgs doublet.
\begin{table}[!ht]%\tab1
\centering
\begin{tabular}{|l|c|l|c|}
%\hline
 \multicolumn{2}{c}{$\psi^2H^4$} &  \multicolumn{2}{c}{ $\psi^2H^3D$}
\\
\hline
$\calO_{LH}$ & $\epsilon_{ij}\epsilon_{mn}(L^iCL^m)H^jH^n(H^\dagger H)$ & $\calO_{LeHD}$ & $\epsilon_{ij}\epsilon_{mn}(L^iC\gamma_\mu e)H^jH^miD^\mu H^n$
\\
\hline
 \multicolumn{2}{c}{$\psi^2H^2D^2$}& \multicolumn{2}{c}{$\psi^2H^2X$}
\\
\hline
$\calO_{LHD1}$ &$\epsilon_{ij}\epsilon_{mn}(L^iCD^\mu L^j)H^m(D_\mu H^n)$ &$\calO_{LHB}$    &$ g_1\epsilon_{ij}\epsilon_{mn}(L^iC\sigma_{\mu\nu}L^m)H^jH^nB^{\mu\nu}$ \\
$\calO_{LHD2}$  & $\epsilon_{im}\epsilon_{jn}(L^iCD^\mu L^j)H^m(D_\mu H^n)$ & $\calO_{LHW}$  &$g_2\epsilon_{ij}(\epsilon \tau^I)_{mn}(L^iC\sigma_{\mu\nu}L^m)H^jH^nW^{I\mu\nu}$ \\
\hline
  \multicolumn{2}{c}{$\psi^4D$}  &   \multicolumn{2}{c}{$\psi^4H$}\\ \hline
$\calO_{\bar{d}uLLD}$ & $\epsilon_{ij}(\bar{d}\gamma_\mu u)(L^iCiD^\mu L^j)$ & $\calO_{\bar{e}LLLH}$ & $\epsilon_{ij}\epsilon_{mn}(\bar{e}L^i)(L^jCL^m)H^n$\\
   &  & $\calO_{\bar{d}LQLH1}$ & $\epsilon_{ij}\epsilon_{mn}(\bar{d}L^i)(Q^jCL^m)H^n$\\
   &  & $\calO_{\bar{d}LQLH2}$ & $\epsilon_{im}\epsilon_{jn}(\bar{d}L^i)(Q^jCL^m)H^n$\\
   &  & $\calO_{\bar{d}LueH}$ & $\epsilon_{ij}(\bar{d}L^i)(uCe)H^j$\\
   &  & $\calO_{\bar{Q}uLLH}$ & $\epsilon_{ij}(\bar{Q}u)(LCL^i)H^j$\\
\hline
\rowcolor{mygray}
$\calO_{\bar{L}QddD}$ & $(\bar{L}\gamma_\mu Q)(dCiD^\mu d)$ & $\calO_{\bar{L}dud\tilde{H}}$ & $(\bar{L}d)(uCd)\tilde{H}$\\
\rowcolor{mygray}
$\calO_{\bar{e}dddD}$  & $(\bar{e}\gamma_\mu d)(dCiD^\mu d)$ & $\calO_{\bar{L}dddH}$ & $(\bar{L}d)(dCd)H$\\
\rowcolor{mygray}
   &  & $\calO_{\bar{e}Qdd\tilde{H}}$ & $\epsilon_{ij}(\bar{e}Q^{i})(dCd)\tilde{H}^j$\\
\rowcolor{mygray}
   &  & $\calO_{\bar{L}dQQ\tilde{H}}$ & $\epsilon_{ij}(\bar{L}d)(QCQ^{i})\tilde{H}^j$ \\ \hline
\end{tabular}
\caption{Dim-7 operators in 6 classes are divided into two subsets with $L=2$ and $B=0$ and $B=-L=1$ (in gray) respectively, where $(D_\mu H^n)$ should be understood as $(D_\mu H)^n$ etc. This table is taken from ref.~\cite{Liao:2016hru}.}
\label{tab4}
\end{table}

%%%%%
\section{RGEs for dim-7 operators in SMEFT relevant to the decay $K^-\rightarrow\pi^+l^-l^-$}
\label{app3}

In section~\ref{sec5} we matched effective interactions between LEFT and SMEFT at the scale $\Lambda_\textrm{EW}$. To connect to new physics at a higher scale $\Lambda_\textrm{NP}$ we have to include RGE effects on the operators in SMEFT. The complete one-loop RGE analysis has been worked out for the subset of operators violating both baryon and lepton numbers in ref.~\cite{Liao:2016hru} and for the subset violating only lepton number in ref.~\cite{Liao:2019tep}. We reproduce here the RGEs for the Wilson coefficients entering the matching conditions equations~\eqref{mat1}-\eqref{mat3}:
\begin{eqnarray}
4\pi\frac{d}{d\ln\mu}C_{\bar{d}uLLD}^{prll\dagger}&=&
\left(\frac{1}{10}\alpha_1
-\frac{1}{2}\alpha_2\right)C_{\bar{d}uLLD}^{prll\dagger},
\\
4\pi\frac{d}{d\ln\mu}C_{LHD2}^{ll\dagger}&=&
\left(\frac{12}{5}\alpha_1
+3\alpha_2+4\alpha_\lambda+6\alpha_t\right)C^{ll\dagger}_{LHD2}
+\left(-8\alpha_2\right)C^{ll\dagger}_{LHD1},
\\
4\pi\frac{d}{d\ln\mu}C_{LHD1}^{ll\dagger}&=&
\left(-\frac{9}{10}\alpha_1
+\frac{11}{2}\alpha_2+6\alpha_t\right)C^{ll\dagger}_{LHD1}
+\left(-\frac{33}{20}\alpha_1-\frac{19}{4} \alpha_2-2\alpha_\lambda \right)C^{ll\dagger}_{LHD2},
\\\label{reglhw}
4\pi\frac{d}{d\ln\mu}C_{LHW}^{ll\dagger}&=&
\left(-\frac{6}{5}\alpha_1+\frac{13}{2}\alpha_2+4\alpha_\lambda
+6\alpha_t\right)C^{ll\dagger}_{LHW}
+\frac{5}{8}\alpha_2C^{ll\dagger}_{LHD1}
+\left(-\frac{9}{80}\alpha_1
+\frac{11}{16}\alpha_2\right)C^{ll\dagger}_{LHD2},
\end{eqnarray}
where $\alpha_i=g_i^2/(4\pi)$ with $g_i$ being the gauge couplings for the SM gauge group $SU(3)_C\times SU(2)_L\times U(1)_Y$, $\alpha_t=y_t^2/(4\pi)$ with $y_t$ being the Yukawa coupling of the top quark (in the convention, $m_t=y_tv/\sqrt{2}$ with $v\approx 246~\GeV$), and $\alpha_\lambda=\lambda/(4\pi)$ with $\lambda$ being the Higgs self-coupling (in the convention, $m_h^2=2\lambda v^2$). The above equations ignore much smaller Yukawa couplings of other fermions.

\end{appendices}

%%%%%

\end{document}